\documentclass[preprint2]{emulateapj}
\usepackage{graphicx}
\usepackage{longtable}
\usepackage{mdwlist}
\usepackage{color}
\usepackage{enumitem}
\usepackage{lipsum}
\slugcomment{version \today: fm}
\shorttitle{Deciphering the large-scale environment of radio galaxies in the local Universe II}
\shortauthors{F. Massaro et al.}

\begin{document}
\title{Deciphering the large-scale environment of radio galaxies in the local Universe II. \\ A statistical analysis of environmental properties}
\author{F. Massaro\altaffilmark{1,2,3,4}, A. Capetti\altaffilmark{2}, A. Paggi\altaffilmark{1,2,3}, R. D. Baldi\altaffilmark{5}, \\ A. Tramacere\altaffilmark{6}, I. Pillitteri\altaffilmark{7}, R. Campana\altaffilmark{8}, A. Jimenez-Gallardo\altaffilmark{1,2} \& V. Missaglia\altaffilmark{9}}
\altaffiltext{1}{Dipartimento di Fisica, Universit\`a degli Studi di Torino, via Pietro Giuria 1, I-10125 Torino, Italy.}
\altaffiltext{2}{INAF-Osservatorio Astrofisico di Torino, via Osservatorio 20, 10025 Pino Torinese, Italy.}
\altaffiltext{3}{Istituto Nazionale di Fisica Nucleare, Sezione di Torino, I- 10125 Torino, Italy.}
\altaffiltext{4}{Consorzio Interuniversitario per la Fisica Spaziale, via Pietro Giuria 1, I-10125 Torino, Italy.}
\altaffiltext{5}{Department of Physics and Astronomy, University of Southampton, Highfield, SO17 1BJ, UK.}
\altaffiltext{6}{University of Geneva, Chemin d'Ecogia 16, Versoix, CH-1290, Switzerland.}
\altaffiltext{7}{INAF-Osservatorio Astronomico di Palermo G.S. Vaiana, Piazza del Parlamento 1, 90134, Italy.}
\altaffiltext{8}{INAF/OAS, via Piero Gobetti 101, I-40129, Bologna, Italy.}
\altaffiltext{9}{Harvard - Smithsonian Astrophysical Observatory, 60 Garden Street, 02138, Cambridge (MA), USA.}

\begin{abstract} 
In our previous analysis we investigated the large-scale environment of two samples of radio galaxies (RGs) in the local Universe (i.e. with redshifts $z_\mathrm{src}\leq$0.15), classified as FR\,I and FR\,II on the basis of their radio morphology. The analysis was carried out using i) extremely homogeneous catalogs and ii) a new method, known as {\it cosmological overdensity}, to investigate their large-scale environments. We concluded that, independently by the shape of their radio extended structure, RGs inhabit galaxy-rich large-scale environments with similar characteristics and richness. In the present work, we first highlight additional advantages of our procedure, that does not suffer cosmological biases and/or artifacts, and then we carry out an additional statistical test to strengthen our previous results. We also investigate properties of RG environments using those of the {\it cosmological neighbors}. We find that large-scale environments of both FR\,Is and FR\,IIs are remarkably similar and independent on the properties of central RG. Finally, we highlight the importance of {\it comparing radio sources in the same redshift bins} to obtain a complete overview of their large-scale environments.
\end{abstract}


\section{Introduction}
\label{sec:intro} 
In the last decades, extensive multifrequency investigations, combined with different statistical procedures, have shown that radio galaxies (RGs) preferentially inhabit galaxy-rich large-scale environments \citep[see e.g.,][]{prestage88,hill91,zirbel97,worrall00,belsole07,tasse08}, making them ideal laboratories to investigate formation and evolution of cosmological structures \citep[see also][]{best04,gendre13,ineson13,ineson15}. Most of these studies compare different classes of RGs, distinguishing between FR\,I (i.e, edge-darkened) and FR\,II (i.e, edge-brightened) classes \citep{fanaroff74} or between low excitation and high excitation RGs \citep[LERGs and HERGS, respectively; see e.g.,][]{hine79,laing94}.

The most efficient approaches, that allows to get a complete overview of both galaxies and intergalactic medium (IGM) are certainly those based or combined with X-ray observations \citep[see e.g.,][]{hardcastle00,ineson13,ineson15}. Although in the last decades, thanks to XMM-{\it Newton} and {\it Chandra} campaigns, X-ray archives were enriched of RG observations \citep[see e.g.,][for recent results]{evans06,croston08,massaro12,mingo14,massaro15,mingo17,massaro18,stuardi18}, the largest fraction of the analyses carried out to date on their large-scale environments are preferentially based on optical and infrared surveys \citep[see e.g.,][]{ching17,miraghei17,massaro19}. This is mainly due to difficulties on performing deep and, at the same time, wide-area X-ray surveys, that would allow us to avoid biases introduced by a scarce sampling of RG populations. Results achieved to date are not always in agreement.

We recently proposed a different approach based on the selection of extremely homogenous datasets, restricted to the local Universe (i.e., source redshifts $z_{src} \leq$0.15), to shed light on RG large-scale environments \citep[][hereinafter M19]{massaro19}. 

We first created two extreme homogeneous catalogs of FR\,I and FR\,II RGs \citep[hereinafter FRICAT and FRIICAT, respectively;][]{capetti17a,capetti17b} at high level of completeness (i.e., $\sim$95\%), based on uniform radio images available thanks to the Faint Images of the Radio Sky at Twenty cm (FIRST) radio survey \citep{white97,helfand15} that has almost the same footprint of the Sloan Digital Sky Survey \citep[see e.g,][]{ahn12} and is also completely covered by the NRAO VLA Sky Survey \citep[NVSS;][]{condon98} and, in the mid-infrared, thanks to observations of the Wide-field Infrared Survey Explorer \citep[i.e., WISE;][]{wright10}. Then our investigation compared the efficiency of several clustering algorithms, including a new procedure, known as {\it cosmological overdensity}, based on the so-called {\it cosmological neighbors} (see the following sections for more details), opportunely developed to avoid cosmological biases and artifacts. 

We mainly concluded that RGs, independently of their radio (FR\,I vs. FR\,II) or optical classification (LERG vs. HERG) tend to inhabit galaxy-rich large-scale environments with similar richness (M19). However, the limited number of HERGs present in our FRIICAT prevented us from drawing firm statistical conclusions when comparing LERG and HERG populations, thus the statement based on the optical classification has to be treated with caution deserving a deeper analysis. Finally, we highlighted the importance of {\it comparing radio sources in the same redshift bins} to obtain a complete overview of their large-scale environments. { It is worth highlighting that searching for galaxy clusters and groups using spectroscopic redshifts is a method already used in the past \citep[see e.g.,][]{danese80,hucra82,girardi93,fadda96}}.

Here we first present additional evidences of cosmological biases and artifacts that could affect analyses of the large-scale environments highlighting the advantages of the cosmological overdensity method. Then we perform an additional statistical test to obtain a direct comparison between the FR\,I and the FR\,II populations. On the basis of the distribution of optical sources surrounding RGs, we also present a new set of parameters to i) characterize and ii) compare their large-scale environments. 

The paper is organized as follows. In \S~\ref{sec:samples} we describe the samples selected to carry out our analysis while in \S~\ref{sec:cn} we provide a brief description of the cosmological overdensity procedure. Additional evidences of cosmological biases and artifacts that could arise in similar analyses is given in \S~\ref{sec:arti}. Then in \S~\ref{sec:flipcoin} we present a new statistical test that strengthens the evidence that FR\,Is and FR\,IIs inhabit similar galaxy-rich large-scale environments, while \S~\ref{sec:environments} is devoted to the description of several ambient parameters obtainable from the distribution of {\it cosmological neighbors}. \S~\ref{sec:xray} is then dedicated on future developments achievable with dedicated X-ray observations. Finally, summary and conclusions are given in \S~\ref{sec:summary} with tables and additional figures reported In Appendix~\ref{app:figtab}. 

As in M19, we adopt cgs units for numerical results and we assume a flat cosmology with $H_0=69.6$ km s$^{-1}$ Mpc$^{-1}$, $\Omega_\mathrm{M}=0.286$ and $\Omega_\mathrm{\Lambda}=0.714$ \citep{bennett14}, unless otherwise stated. Thus, according to these cosmological parameters, 1\arcsec\ corresponds to 0.408 kpc at $z_\mathrm{src}=$0.02 and to 2.634 kpc at $z_\mathrm{src}=$0.15.

\begin{figure*}[!th]
\begin{center}
\includegraphics[height=6.2cm,width=8.4cm,angle=0]{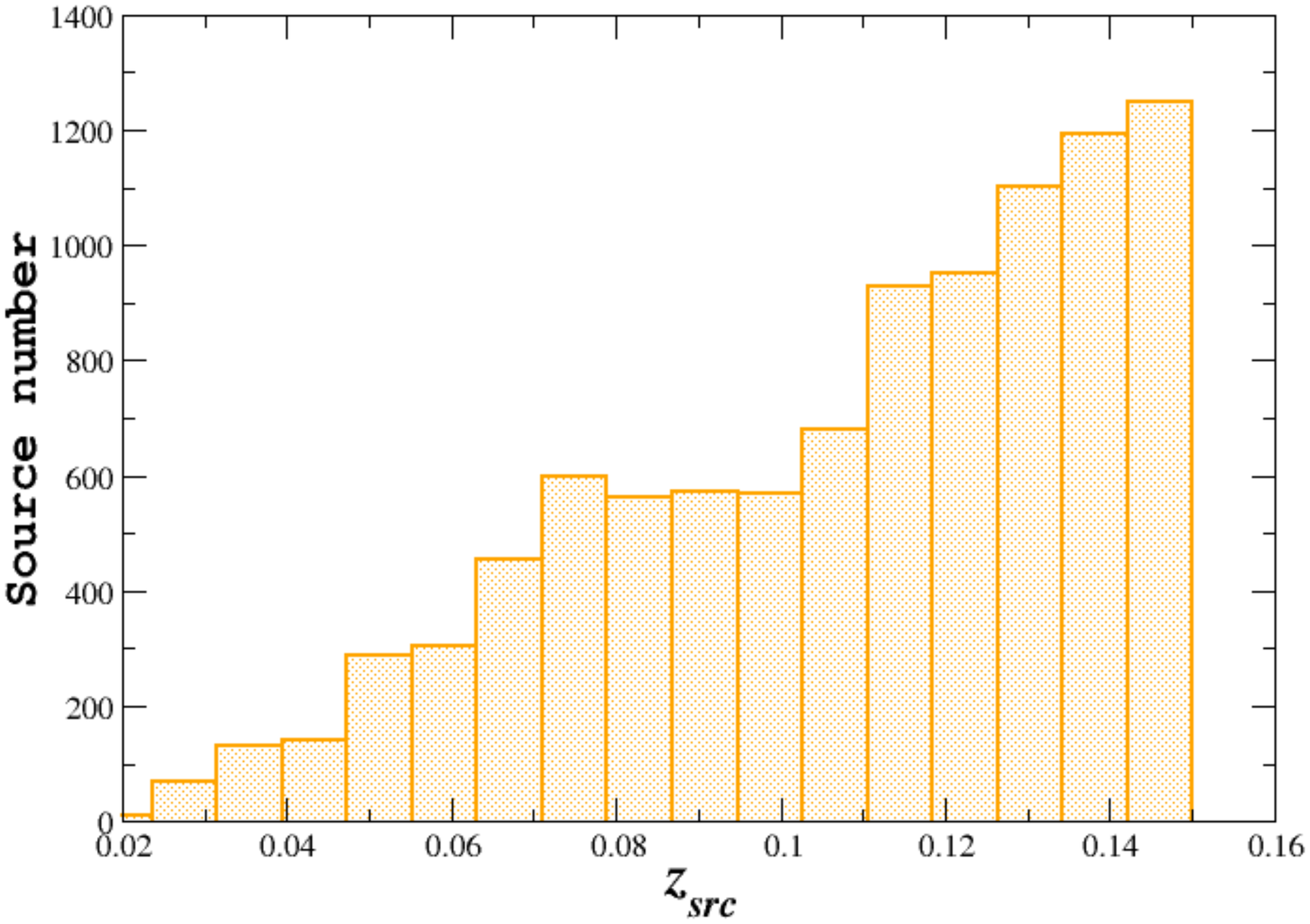}
\includegraphics[height=6.2cm,width=8.4cm,angle=0]{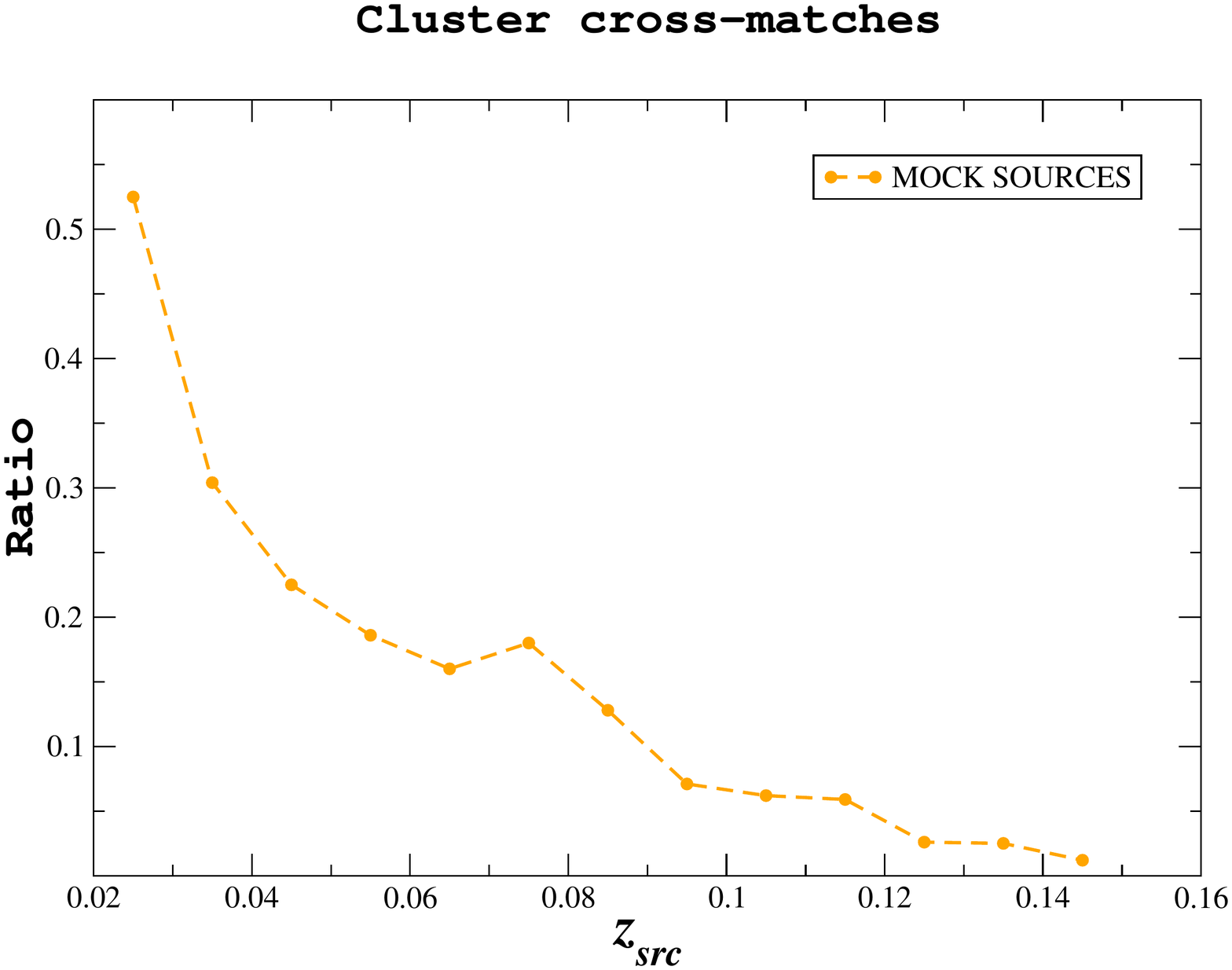}
\end{center}
\caption{Left panel) The $z_{src}$ distribution of 9800 random positions (i.e., mock sources) chosen in the SDSS footprint, lacking a radio counterpart within 5\arcsec. Right panel) The ratio between the number of mock sources having a group or a cluster of galaxies with richness $N_{gal}>4$ from the T12 catalog and with having $\Delta\,z = |z_\mathrm{src}-z_\mathrm{cl}|\leq$0.005, and the total number of mock sources in bin of $z_{src}$ of size 0.01. This confirms our previous results based on a mock catalog of 5000 sources \citep{massaro19} that it is easier to find a positive cross-match at low $z_{src}$.}
\label{fig:mockz}
\end{figure*}

\section{Sample selection}
\label{sec:samples} 
We selected two RG catalogs to carry out our analysis starting from the radio-loud sample of Best \& Heckman (2012).

The first catalog is the combination of FRICAT and sFRICAT both described in Capetti et al. (2017a). The FRICAT sources, chosen on the basis of their FR\,I radio morphology, are selected to have a radio structure beyond a distance of 30 kpc, measured from the optical position of the host galaxy, with the only exceptions of the 14 sFRICAT objects with radio extended emission having the same morphology but extending between 10 and 30 kpc but limited to $z_{src}=$0.05 \citep[see][for details]{capetti17a}. This combination of these two samples includes RGs at redshift $z_{src} \leq 0.15$, all hosted in red early-type galaxies and spectroscopically classified low excitation radio galaxies (LERGs) for a total of 209 radio sources. 

The second sample is the FRIICAT \citep{capetti17b}, composed of 105 edge-brightened radio sources (FR\,II type) within the same redshift range of the previous catalog. About $\sim$90\% of the FR\,IIs listed in the FRIICAT are spectroscopically classified as LERGs, being hosted red early-type galaxies as occurs for FR\,Is. The remaining $\sim$10\% of the FRIICAT shows indeed optical spectra typical of high excitation radio galaxies (HERGs) and host galaxies bluer in the optical band and redder in the infrared than FR\,II LERGs.

Thanks to their selection criteria both RG catalogs are not contaminated by compact radio objects, as compact steep spectrum sources and FR\,0s \citep{baldi15,baldi18}, which show a different cosmological evolution, with respect to FR\,Is and FR\,IIs and could potentially lie in different environments.

The selected RG catalogs include only sources lying in the central footprint of the SDSS, because it is the same sky area covered by the main catalog of groups and clusters of galaxies adopted in our analysis: the one created by Tempel et al. (2012, hereinafter T12). The T12 catalog has the largest number of cluster/group detections with spectroscopic redshifts $z_\mathrm{cl}$ in the range between 0.009 and 0.20 and peaking around 0.08. This catalog of groups and clusters is based on a modified version of the Friends-of-Friends algorithm \citep{hucra82,tago10} and its richness is indicated as $N_{gal}$.

\section{Cosmological overdensity}
\label{sec:cn} 
Our analysis on the large-scale environment of RGs is based on the definition of two type of optical sources lying within 2\,Mpc. These are the i) cosmological neighbors and the ii) candidate elliptical galaxies whose definition is reported below, for sake of completeness.

\begin{enumerate}
\item {\it Cosmological neighbors:} all optical sources lying within the 2\,Mpc radius computed at $z_\mathrm{src}$ of the central radio galaxy with all the SDSS magnitude flags indicating a galaxy-type object (i.e., {\it uc=rc=gc=ic=zc=}3), and having a spectroscopic redshift $z$ with $\Delta\,z=|z_\mathrm{src}-z|\leq$0.005, corresponding to the maximum velocity dispersion in groups and clusters of galaxies \citep[see e.g.,][]{moore93,eke04,berlind06}. 

\suspend{enumerate}
We indicate the number of cosmological neighbors, $N_{cn}^{500}$, $N_{cn}^{1000}$ and $N_{cn}^{2000}$, lying within 500\,kpc, 1\,Mpc and 2\,Mpc distance from the central radio galaxy, respectively, that provide an estimate of the environmental richness. 
\resume{enumerate}

\item {\it Candidate elliptical galaxies:} all SDSS sources, lying within the 2\,Mpc distance from the central RG at $z_\mathrm{src}$ and with $u-r$ and $g-z$ optical colors consistent with those of a sample of elliptical galaxies, chosen out of the galaxy zoo project \citep{lintott08}, at the same redshift and within the 90\% level of confidence evaluated using the Kernel Density Estimator \citep{richards04,dabrusco09,massaro11,massaro13}. 
\end{enumerate}
\noindent
We selected elliptical-type galaxies because their density in galaxy groups or clusters is larger than that of spirals \citep[see e.g.,][]{biviano00}.

Source selected as candidate elliptical galaxies do not necessarily have spectroscopic $z_\mathrm{src}$ estimates. 

Candidate elliptical galaxies were mainly used in the previous analysis with the following two aims. First, they allow to check galaxy rich environments also when the number of cosmological neighbors is limited, since we were able to test the presence of the ``red sequence'' \citep[i.e., the well-known relation between colors and magnitude for galaxies that are members of groups and/or clusters][]{visvanathan77,gladders98,gladders00} with two different color magnitude diagrams (see M19 for more details). Then the distance between the 5$^{th}$ candidate elliptical galaxy and the central RG was used to compute the $\Sigma_5$ parameter as an estimator of the richness \citep[see e.g.,][]{sabater13,worpel13,haas12}.

Based on our criterion, a cosmological overdensity occurs when the number of cosmological neighbors counted within 500 kpc is higher than the 95\% quantile of the $N_{cn}^{500}$ distribution measured for a sample of mock sources located in random positions of the sky. This $N_{cn}^{500}$ threshold has been chosen in bin of $z_\mathrm{src}$ of size 0.01 to compare sources at same redshifts thus to avoid cosmological biases and artifacts. At redshifts above 0.1 we imposed to have at least two cosmological neighbors within 1\,Mpc in addition to the previous criterion on $N_{cn}^{500}$. Additional and more specific details about this procedure can be found in M19.

\begin{figure}[!b]
\begin{center}
\includegraphics[height=6.6cm,width=8.4cm,angle=0]{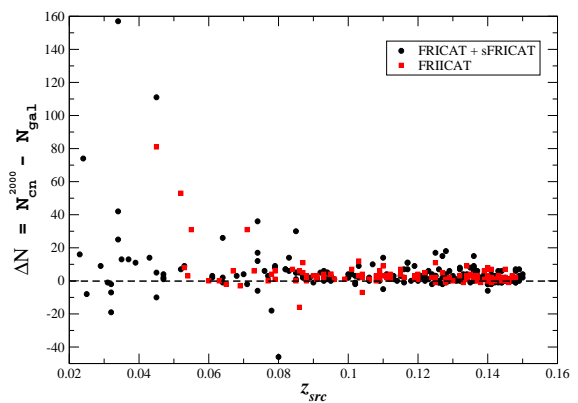}
\end{center}
\caption{The difference $\Delta\,N$ between the number of cosmological neighbors, $N_{cn}^{2000}$ lying within 2\,Mpc distance from the central RG and $N_{gal}$ the richness estimated in the T12 cluster catalog. Black circles mark radio sources in the FRICAT, including those of the sFRICAT, while red squares correspond to the FR\,IIs.}
\label{fig:dN}
\end{figure}

\section{Cosmological biases and artifacts}
\label{sec:arti} 
Here we describe several cosmological biases and artifacts encountered while carrying out our investigation to highlight advantages of using, whenever possible, the method of cosmological overdensity.

\subsection{Biases in the catalog crossmatches}
\label{sec:cross} 
Cross-matches with catalogs of groups and clusters of galaxies shows a clear $z_{src}$ dependence. As shown in Fig.~\ref{fig:mockz}, we built a sample of 9800 random positions in the SDSS footprint and cross-matched it with the T12 catalog. The redshift distribution of mock sources rises up to $z_{src}=$0.15 as those of FRICAT and FRIICAT.

The fraction of mock sources per redshift bin of size 0.01 having a positive crossmatch with a group or a cluster of galaxies within 2\,Mpc, computed a its $z_{src}$, and having $\Delta\,z = |z_\mathrm{src}-z_\mathrm{cl}|\leq$0.005 and $N_{gal}>4$, significantly decreases while $z_{src}$ increases. This indicates that we could expect a higher chance of finding a low redshift RG associated with a group or a galaxy cluster than a high redshift one. Thus we chose the thresholds on the number of cosmological neighbors in the cosmological overdensity as function of $z_{src}$.

A second problem arises when using catalog crossmatches regarding the richness estimated. In Fig.~\ref{fig:dN} we show the difference $\Delta\,N$ between the number of cosmological neighbors, $N_{cn}^{2000}$ lying within 2\,Mpc distance from the central RG and $N_{gal}$ from the T12 group/cluster catalog. It is quite evident that the $N_{gal}$ parameter underestimates the group/cluster richness. There are only a few cases where $N_{cn}^{2000}$ provides a lower estimate of the group/cluster richness. However,  these cases could be also due to an incorrect run of the algorithm of the T12 catalog since we did not detect sources lying at similar redshift of the central RG and not even candidate elliptical galaxies for most of them. It is worth noting that the richness estimated in cluster catalogs is generally decreasing with redshift, as for example occurs for $N_{gal}$ in the T12, thus expecting to find less galaxy-rich environments as $z_{src}$ increases.
 
\begin{figure}[!t]
\begin{center}
\includegraphics[height=6.6cm,width=8.4cm,angle=0]{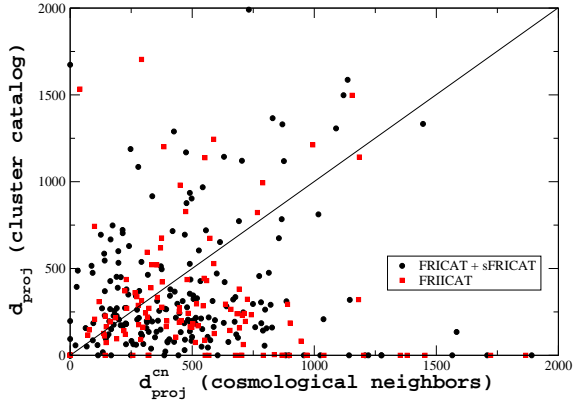}
\end{center}
\caption{On the x axis we report the projected distance $d_{proj}$ measured between the position of the RG in the center of the field examined and the centroid of the spatial distribution of cosmological neighbors within 2\,Mpc while on the y axis the projected distance is computed using the location of the closest group/cluster positionally associated with the T12 catalog. As in Fig.~\ref{fig:dN} black circles mark radio sources in the FRICAT, including those of the sFRICAT, while red squares correspond to the FR\,IIs.}
\label{fig:dproj}
\end{figure}

Additional advantages of the method based on the cosmological overdensity with respect to the cluster catalog crossmatches regard the estimate of the position of the central RG with respect to cluster center both in angular separation and redshift. In Fig.~\ref{fig:dproj} we show the projected distance $d_{proj}$ measured between the position of the RG in the center of the field examined and i) the centroid of the spatial distribution of cosmological neighbors within 2\,Mpc (x-axis) or ii) that of the group/cluster reported in the T12 catalog (y-axis). The majority of sources have values below bisector indicating that $d_{proj}$ estimated via cosmological neighbors tend to be smaller than those evaluated with the T12 catalog.

\begin{figure}[!h]
\begin{center}
\includegraphics[height=6.6cm,width=8.4cm,angle=0]{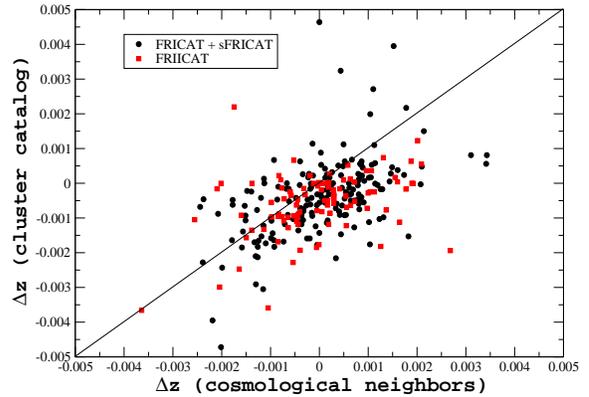}
\end{center}
\caption{The redshift difference $\Delta\,z$ between the central RG $z_{src}$ and the average value of the of cosmological neighbors lying within 2\,Mpc (x-axis) in comparison with that estimated using the spectroscopic redshift of the closest cluster associated with the T12 catalog (y-axis). Sources belonging to both FRICAT and sFRICAT are shown as black circles while FR\,IIs as red squares.}
\label{fig:dz}
\end{figure}

Finally, in Fig.~\ref{fig:dz} we show the redshift difference $\Delta\,z$ between the central RG $z_{src}$ and i) the average value of the of cosmological neighbors lying within 2\,Mpc (x-axis) and ii) the redshift of the closest cluster associated using the T12 catalog (y-axis). Once again the largest fraction of RG in both FRICAT and FRIICAT lies below the bisector, highlighting that cosmological neighbors provide a better sampling of the RG large-scale environment. Assuming that all RGs belong to a galaxy group/cluster, thanks to the distribution of cosmological neighbors we can achieve a better estimate of its redshift: $z_\mathrm{cl}$.

\subsection{Cosmological dependences of noise}
\label{sec:noise} 
\begin{figure}[!th]
\begin{center}
\includegraphics[height=6.2cm,width=8.4cm,angle=0]{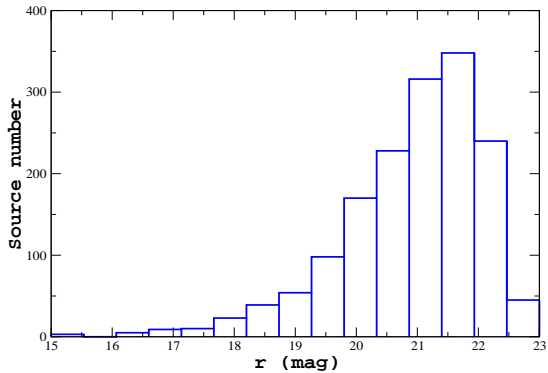}
\end{center}
\caption{The distribution of the $R$ band apparent magnitude $r$ of all SDSS sources classified as galaxies (i.e., {\it uc=rc=gc=ic=zc=}3) in the 2\,Mpc field of the FR\,I SDSSJ073505.25+415827.5. It is evident how the number of sources increases drastically when $r$ increases up to the completeness level of the SDSS.}
\label{fig:mag}
\end{figure}

Selecting candidate elliptical galaxies on the basis of their spectroscopic redshift, whenever possible, is more precise than for example performing a simple magnitude cut as to estimate the Abell environmental richness. Counting overdensity of optical sources within a certain distance in kpc and with apparent magnitude $m\leq\,m_{src}+2$, with $m_{src}$ being the apparent magnitude of the central radio source \citep[see e.g.,][]{hill91}, introduces a significant amount of { noise} due to background and foreground objects being counted. This occurs because the number of optical galaxies increases significantly with the magnitude as shown for example in Fig.~\ref{fig:mag} for the $R$ band apparent magnitude of galaxy-type objects (i.e., {\it uc=rc=gc=ic=zc=}3) in the 2\,Mpc field of the FR\,I SDSSJ073505.25+415827.5. 
Thus comparing overdensity of optical sources around RG lying at different $z_{src}$ when having the same absolute magnitude, or simply at different apparent magnitudes, implies comparing sources with a different signal-to-noise ratio. This makes the comparison between source populations having a different redshift distribution and/or different apparent magnitude distribution and/or a similar intrinsic luminosity/absolute magnitude, as FR\,Is and FR\,IIs, challenging. This effect could be actually mitigated by counting number sources per unit of area, since at larger redshifts areas of same physical size appear smaller thus containing less background/foreground sources (i.e., less noise), but, on the other hand, it does not help since it introduces a cosmological dependence of the noise.

\begin{figure}[!t]
\begin{center}
\includegraphics[height=6.2cm,width=8.4cm,angle=0]{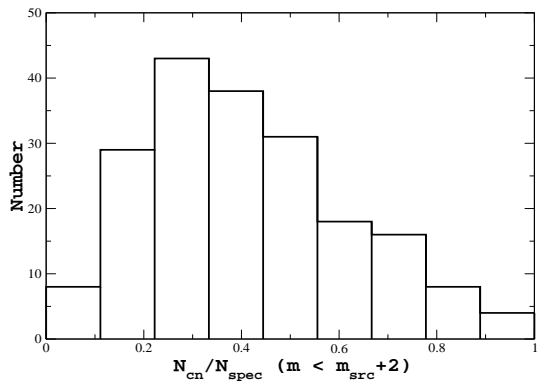}
\end{center}
\caption{The distribution of the ratio between the total number of cosmological neighbors and that of optical sources with spectroscopic redshifts both selected to have $m\leq\,m_{src}+2$, in the $R$ band, in the field of all FR\,Is lying within 2\,Mpc form the central RG. This distribution shows how selecting sources only on the basis of their SDSS magnitude introduce { more noise} since out of those with a spectroscopic redshift only a tiny fraction, on average, appear to be gravitationally bounded with the central RG.}
\label{fig:fraction}
\end{figure}

In addition, sources selected to be galaxies above a certain threshold of apparent magnitude do not necessarily belong to the same group or cluster of galaxies. In Fig.~\ref{fig:fraction}, we show the ratio between the total number of cosmological neighbors and that of optical sources with spectroscopic redshifts both selected to have $m\leq\,m_{src}+2$ in the $R$ band, similar to the definition of Abell classes \citep[see e.g.,][]{abell89}, within 2\,Mpc distance from all FR\,Is listed in the FRICAT. Only 30-40\% are generally lying at similar redshifts being so classifiable as cosmological neighbors, while others are simply background and foreground sources (i.e., noise), with very different redshifts, and thus erroneously counted. 
Noise-related effects related to the source selection with certain apparent magnitude can be partially mitigated by using photometric redshift estimates, as those provided in the SDSS, and counting sources on the basis of their absolute magnitudes and $V-R$ color \citep[see e.g.,][and references therein]{blanton00,blanton01,wing11}, but not fully avoided. On the other hand, as shown in M19, the cosmological overdensity method does not suffer this noise effect since the threshold to indicate sources in a galaxy-rich large-scale environment is chosen as function of redshifts and because we compare RGs in the same redshift bin.

\begin{figure*}[!t]
\begin{center}
\includegraphics[height=5.6cm,width=6cm,angle=0]{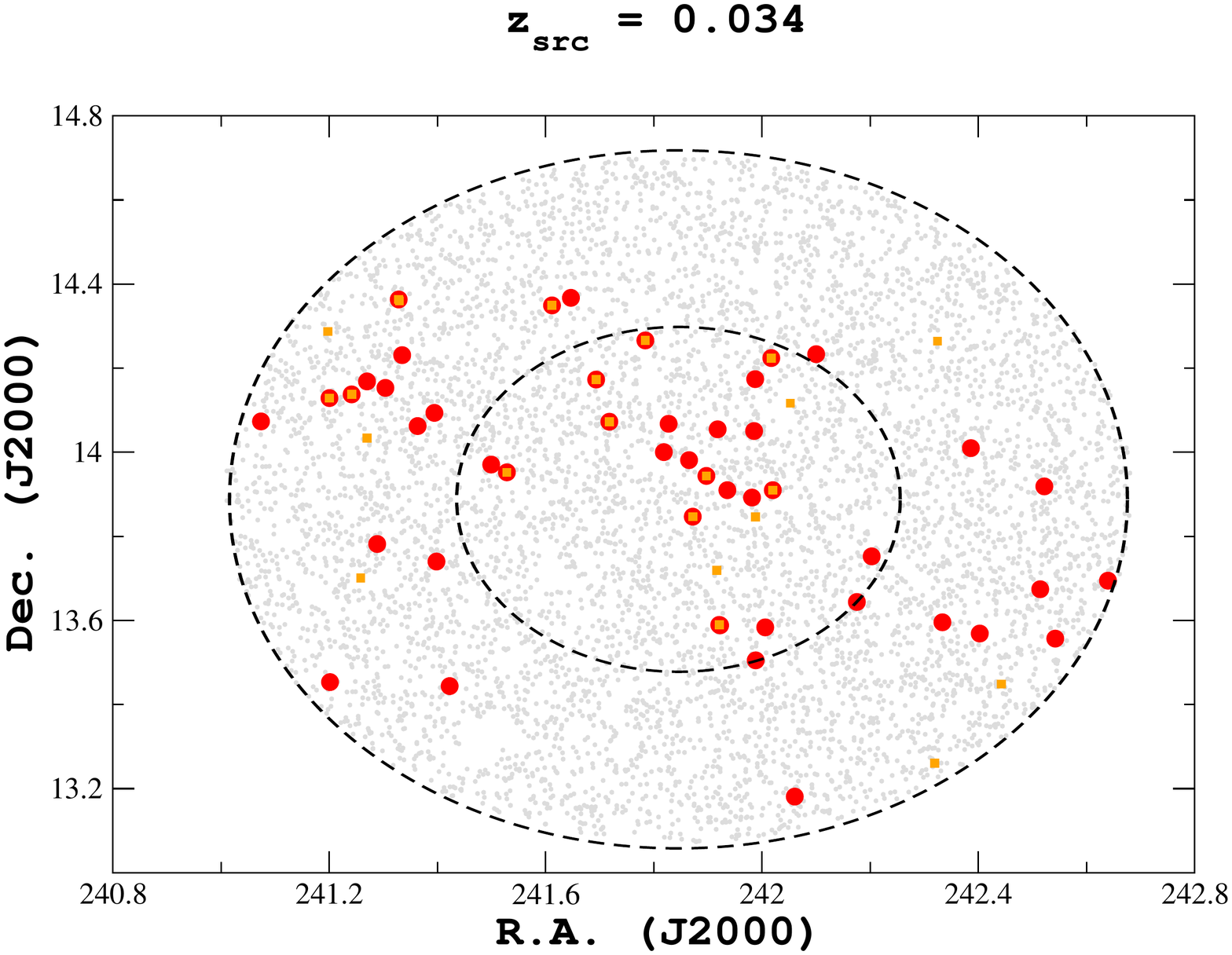}
\includegraphics[height=5.6cm,width=6cm,angle=0]{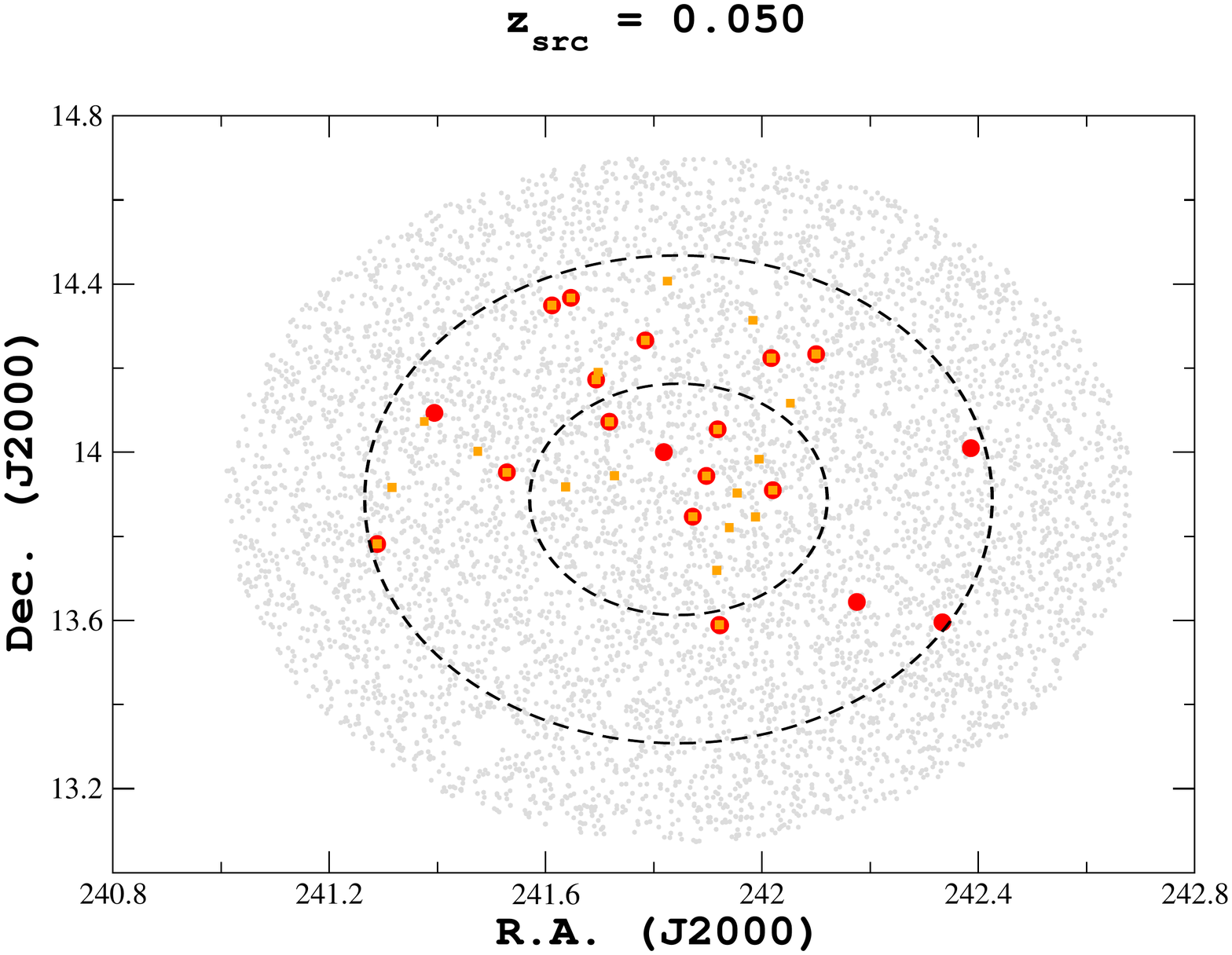}
\includegraphics[height=5.6cm,width=6cm,angle=0]{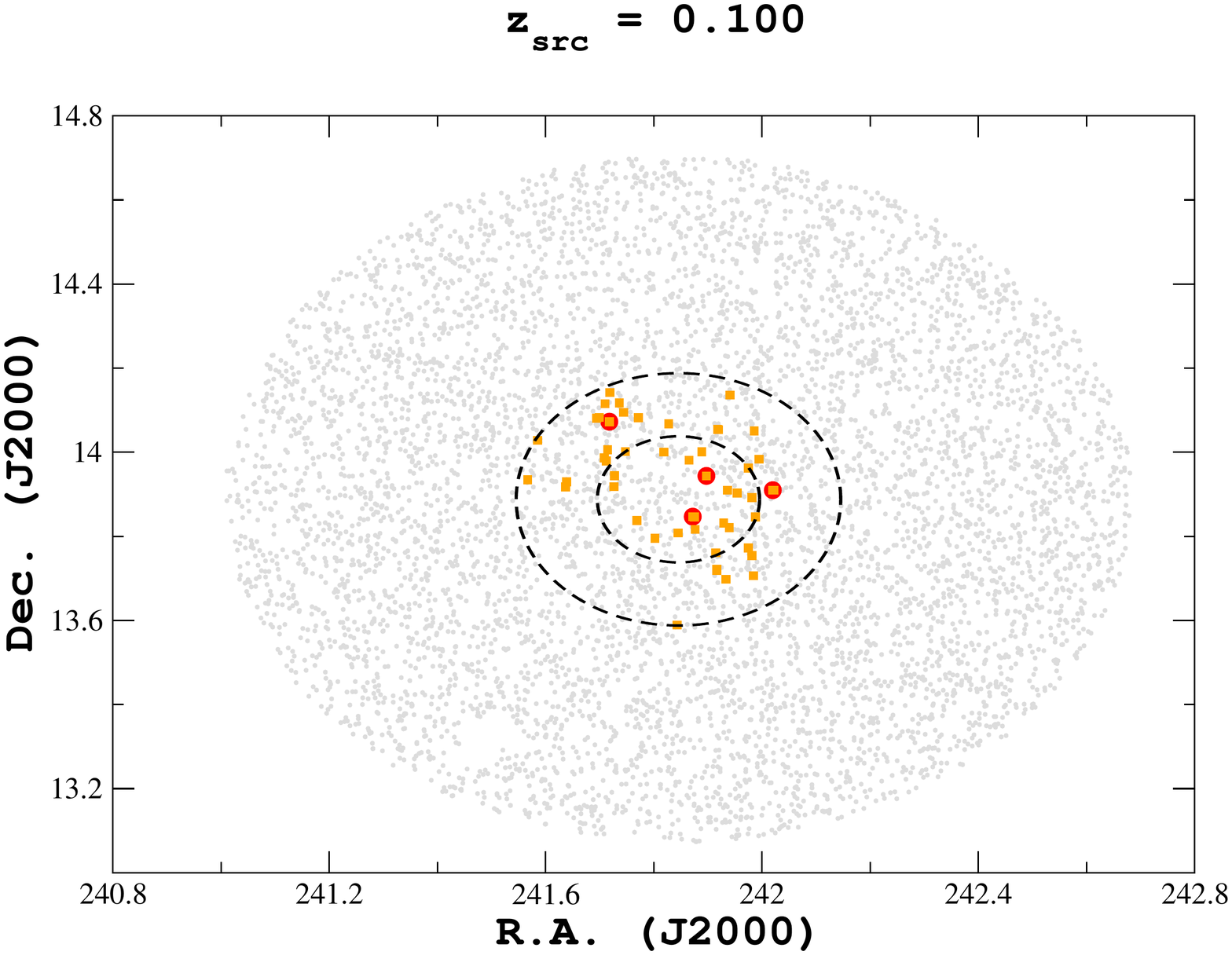}
\includegraphics[height=5.6cm,width=6cm,angle=0]{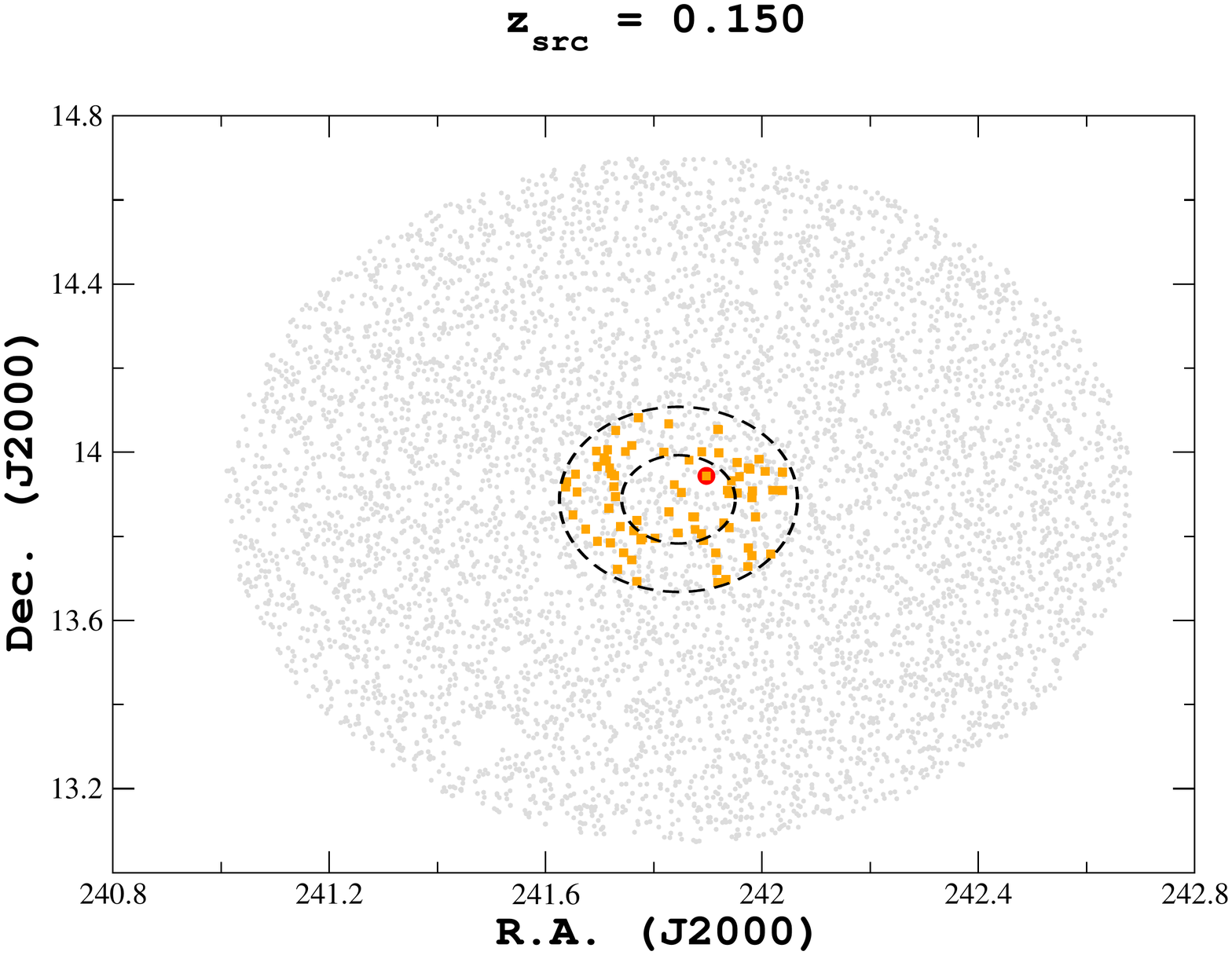}
\end{center}
\caption{The spatial distribution of SDSS optical sources lying within 2\,Mpc from the central RG at $z_{src}=$0.034 and having apparent magnitude in the $R$ band greater than 19 shown as grey circles. Cosmological neighbors highlighted as red circles. Orange squares mark the location of optical galaxies with $m\leq\,m_{src}+2$, to highlight how the noise increases moving at higher redshift. Dashed black lines refer to 1\,Mpc and 2\,Mpc distance from the central RG, respectively. In all other panels we show the distribution of cosmological neighbors computed assuming their intrinsic absolute magnitude $M_r$ and rescaling their apparent magnitudes $m_r$ at different redshifts, namely 0.05 (top right), 0.10 (bottom left) and 0.15 (bottom right), respectively. To create all plots at different redshifts than 0.034 we applied a cut for the rescaled apparent magnitude of the cosmological neighbors selecting only those brighter than 17.8. This limit corresponds to the SDSS criterion to select targets for spectroscopic observations. The number of cosmological neighbors decreasing when changing their distances maintaining the intrinsic power of their host galaxies strengthens our choice of having different thresholds at different redshifts to estimate cosmological overdensity.}
\label{fig:evol}
\end{figure*}

\subsection{Redshift evolution of the cosmological overdensity}
\label{sec:evol} 
Here we show the underlying reason of adopting an ``adaptive'' procedure, as the cosmological overdensity method, with a threshold on the number of cosmological neighbors changing with redshift $z_{src}$. 

In Fig~\ref{fig:evol} we show the 2\,Mpc field around the FR\,I SDSS J160722.95+135316.4 at $z_{src}=$0.034, where red circles mark the locations of cosmological neighbors and dashed lines the distances of 1\,Mpc and 2\,Mpc, respectively. We computed the absolute magnitude in the $R$ band of all cosmological neighbors and then maintaining their intrinsic power we rescaled them at larger distances, namely at redshifts 0.05, 0.10 and 0.15. We also recomputed the radii of the circles of 500\,kpc, 1\,Mpc and 2\,Mpc at each redshift.

In each plot of Fig~\ref{fig:evol} we only reported those cosmological neighbors having the rescaled apparent magnitude $m_r$ brighter than 17.8 corresponding to the SDSS criterion to select targets for spectroscopic observations. For the case of SDSS J160722.95+135316.4 the number of cosmological neighbors within 500\,kpc (i.e., $N_{cn}^{500}$) decreases from 10 at $z_{src}=$0.034, to 3, 2 at $z_{src}=$0.05, 0.10 and none at $z_{src}=$0.15, resepctively, while $N_{cn}^{2000}$ is actually 47 at 0.034 and 20, 4, 1 at $z_{src}=$0.05, 0.10, 0.15, respectively. 

This suggests that the threshold chosen to indicate a galaxy-rich large-scale environment must be redshift dependent as we defined in our previous analysis on the basis of the MonteCarlo simulations performed on a large sample of random positions in the sky. We highlight that the above calculation does not take into account of any cosmological evolution of the groups or the clusters where RGs lie, it is a pure effect of the SDSS selection of spectroscopic targets that decreases the number of cosmological neighbors when the redshift increases as shown in our previous analysis \citep{massaro19}.

Finally, we also show in Fig~\ref{fig:evol} the number of optical galaxies with $m\leq\,m_{src}+2$, to highlight how the noise increases moving at higher redshift where the absolute magnitude of the central RG is fixed and it apparent magnitude $m_{src}$ is rescaled.

\section{Statistical analyses}
\label{sec:flipcoin} 
To provide an additional evidence that the both FR\,Is and FR\,IIs live in galaxy-rich large-scale environment, we performed the following statistical tests all based on the number of cosmological neighbors $N_{cn}$ surrounding them. It is worth noting that, as previously stated, the FRICAT used here is larger than the one adopted in our previous analysis since we also included the sFRICAT sample, allowing us to get a better sampling of the FR\,I population at lower redshifts.

The first test is based on the median of both FR\,I and FR\,II $N_{cn}$ distributions. We computed median values of $N_{cn}^{500}$ for FR\,Is (i.e., $\bar{N}_{cn}^{500}$), per bin of $z_{src}$ of size 0.01, and then we compared the distribution of $N_{cn}^{500}$ for FR\,IIs with $\bar{N}_{cn}^{500}$. We expect that if both RG populations live in large-scale environment of similar richness, 50\% of the FR\,IIs would lie above the median value computed for the FR\,Is while 50\% below, for each bin of $z_{src}$. We also applied the same test to $N_{cn}^{2000}$. All medians values of both $N_{cn}^{500}$ and $N_{cn}^{2000}$ computed for the FR\,I population are reported in Table~\ref{tab:medians}, together with the number of FR\,IIs that lie above and below the medians or having the same number of cosmological neighbors, per redshift bin. 

\begin{figure}[!ht]
\begin{center}
\includegraphics[height=7.2cm,width=9.2cm,angle=0]{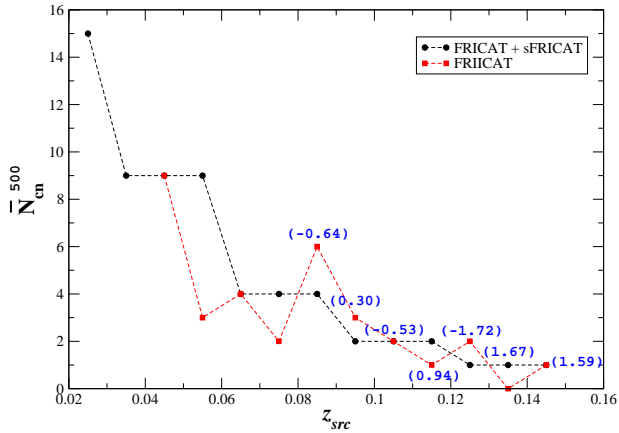}
\end{center}
\caption{The median values for the FR\,I (black) and the FR\,II (red) RGs per bin of redshift $z_{src}$. Together with the median measurements of FR\,Is we also report the number of FR\,IIs that lie above and below them in red and for each $z_{src}$ bin. As shown these fractions of FR\,IIs are almost equally distributed around the medians of the FR\,I population as expected for sources inhabiting similar large-scale environment, independently of their radio extended structure. Blue numbers reported in parenthesis above each median value, at all redshift bins above 0.08, correspond to values computed for the $z_U$ normalized variable of the Mann-Whitney $U$ rank test (see \S.~\ref{sec:flipcoin} for more details).}
\label{fig:medians}
\end{figure}

For $N_{cn}^{500}$, in all 11 $z_{src}$ bins explored, where we can compare FR\,Is and FR\,IIs, we found that in 4 cases the FR\,II median values are greater than $\bar{N}_{cn}^{500}$ computed for FR\,Is while in other 3 cases the situation is the opposite, and in all remaining cases medians of the two populations show the same medians. This is in agreement with the expectations of a binomial distribution, thus strengthening our results that, independently of their radio morphology, both classes of FR\,Is and FR\,IIs live in similar galaxy-rich large-scale environment. The median values $\bar{N}_{cn}^{500}$ for both population of FR\,Is and FR\,IIs are shown in Fig.~\ref{fig:medians}. Then comparing $\bar{N}_{cn}^{2000}$ the situation is in agreement with previous results with 5 cases for which $\bar{N}_{cn}^{2000}$ of FR\,IIs is lower than that for the FR\,I population and 5 cases for which it is greater. Consequently, we can firmly confirm that the large-scale environments of the two classes of RGs, in each considered redshift range, is consistent. 

Unfortunately, this first statistical test does not consider the two populations symmetrically, and it throws away almost all information about the distribution of environments present in the data. For example, if FR\,IIs has a systematically broader distribution of $N_{cn}$ than FR\,Is, or a small tail of very rich, or very poor, environments, this first statistical test would be not sensitive to it. Thus, as second statistical check, we also applied the Mann-Whitney $U$ rank test (a.k.a. Wilcoxon-Mann-Whitney test) to each redshift bin. In Fig.~\ref{fig:medians} we also report, above each value of the medians of the two RG distributions of $N_{cn}$ the normalized $z_U$ variable, defined as $\frac{U-E(U)}{\sigma_U}$), computed for the Mann-Whitney $U$ test and this is always consistent with zero, as expected, within 2 $\sigma$ level of confidence, with the only exception of a single bin between $z=0.13$ and $z=0.14$ where the number $N_{cn}$ for the FR\,IIs appears systematically lower than the median of FR\,Is (see Table~\ref{tab:medians}). It is worth noting that we did not apply this test at $z<$0.08 given the low number of FR\,IIs in these bins. Moreover, also this second statistical test is in agreement with our claim that the large-scale environments of FR\,Is and FR\,IIs cannot be distinguishable. 

Then, for all redshift bins above 0.08 we also performed the two-sided Kolmogorov-Smirnov test between the distributions of $N^{500}_{cn}$ and $N^{2000}_{cn}$ of both FR\,Is and FR\,IIs, results on the $D$ variable compute as well as correspondent $p$-values are reported in Table~\ref{tab:ks}, where in agreement with previous tests and { the null hypothesis that the two distributions are the same cannot be rejected.}

\begin{table} 
\caption{Comparison between $\bar{N}_{cn}^{500}$ and $\bar{N}_{cn}^{2000}$ of FR\,Is with FR\,IIs}
\label{tab:medians}
\begin{center}
\begin{tabular}{lrrrr}
\hline
$z_{src}$ & $\bar{N}^{500}_{cn}$ & $N_{FRII}$ & $\bar{N}^{2000}_{cn}$ & $N_{FRII}$ \\ 
 & & $> \,< \, =$ & & $> \, < \, =$ \\ 
\noalign{\smallskip}
\hline 
\noalign{\smallskip}
0.045 & 9 & 0 - 0 - 1 & 44 & 1 - 0 - 0 \\  
0.055 & 9 & 1 - 4 - 0 & 50 & 1 - 4 - 0 \\  
0.065 & 4 & 1 - 2 - 1 & 18 & 2 - 1 - 1 \\  
0.075 & 4 & 2 - 4 - 0 & 23 & 2 - 4 - 0 \\ 
0.085 & 4 & 6 - 2 - 0 & 12 & 6 - 1 - 1 \\ 
0.095 & 2 & 5 - 4 - 0 & 11 & 3 - 6 - 0 \\ 
0.105 & 2 & 5 - 5 - 4 &  9 & 6 - 7 - 1 \\ 
0.115 & 2 & 2 - 7 - 3 &  5 & 6 - 5 - 1 \\ 
0.125 & 1 & 6 - 0 - 5 &  5 & 6 - 5 - 0 \\ 
0.135 & 1 & 3 -10 - 4 &  8 & 1 -15 - 1 \\ 
0.145 & 1 & 3 -14 -11 &  3 &13 -13 - 2 \\ 
\noalign{\smallskip}
\hline
total & - & 34 - 52 - 29 & - & 47 - 61 - 7 \\
\noalign{\smallskip}
\hline
\end{tabular}\\
\end{center}
Note: col. (1) report the central value of each redshift bin while col. (2) and col. (4) median values for the $N^{500}_{cn}$ and $N^{2000}_{cn}$ parameters computed for the FR\,I population and in comparison with the number of FR\,IIs (i.e., $N_{FRII}$) lying above ($>$), below ($<$) these median values of having being the same ($=$) as shown in col. (3) and col. (5), respectively.\\
\end{table}

Finally, we verified that the shape of the two $N_{cn}$ distributions with a third statistical check computing the skewness and the Pearson median skewness and verifying that the sign of both these parameters corresponds. This occurs in all redshift bins above 0.1 while below, due to the poor number of FR\,II radio galaxies in each $z$ bin strongly affects both parameters providing not significant results. Again this last test provide results in agreement with the previous ones.

\begin{table} 
\caption{Results of the two-sided Kolmogorov-Smirnov test run for the FR\,I and the FR\,II distributions of $N^{500}_{cn}$ and $N^{2000}_{cn}$.}
\label{tab:ks}
\begin{center}
\begin{tabular}{lrrrr}
\hline
$z_{src}$ & D($N^{500}_{cn}$) & p($N^{500}_{cn}$) & D($N^{2000}_{cn}$) & p($N^{2000}_{cn}$) \\ 
\noalign{\smallskip}
\hline 
\noalign{\smallskip}
0.085 & 0.289 & 0.804 & 0.337 & 0.629   \\
0.095 & 0.291  & 0.760 & 0.368 & 0.469  \\ 
0.105 & 0.232 & 0.816 & 0.223 & 0.851  \\
0.115 & 0.172 & 0.986 & 0.235 & 0.831  \\
0.125 & 0.296  & 0.499 & 0.269 & 0.622   \\
0.135 & 0.224 & 0.650 & 0.382 & 0.084 \\
0.145 & 0.205 & 0.445 & 0.089 & 0.999 \\
\noalign{\smallskip}
\hline
\end{tabular}\\
\end{center}
Note: col. (1) report the central value of each redshift bin, while col. (2) and col. (4) the $D$ statistical variable and col. (3) and col.(5) the correspondent $p$-value for the distribution of $N^{500}_{cn}$ and $N^{2000}_{cn}$, respectively.\\
\end{table}

\section{Environmental properties}
\label{sec:environments} 

\subsection{Environmental parameters: definition}
\label{sec:parameters} 
On the basis of the distribution of the cosmological neighbors it is possible to estimate several parameters to investigate the properties of the large-scale environments surrounding FR\,Is and FR\,IIs and compare them with those of the central  RGs.  

In particular, we defined the following quantities.
\begin{itemize}
\item The concentration parameter $\zeta_{cn}$. It is defined as the ratio of the number of cosmological neighbors lying within 500\,kpc and those within 1\,Mpc. This parameter allows us to test if the RG tends to lie close to the center of the group or clusters of galaxies around it or in its outskirt. A similar concentration parameter has been estimate using candidate elliptical galaxies rather than cosmological neighbors: $\zeta_{el}$.
\item The average projected distance $d_m^{cn}$ of the distribution of cosmological neighbors computed between their position and that of the central RG. 
\item The standard deviation $\sigma_{z}$ of the redshift distribution of the cosmological neighbors surrounding each RG.
\end{itemize}
All values for the environmental parameters described above are also reported in Table~\ref{tab:main} in Appendix.

\subsection{Environmental parameters: statistical analysis}
\label{sec:statistics} 
We investigated the distribution of the environmental parameters previously defined searching for possible trends with the RG properties, such as the absolute magnitude in the $R$ band $M_r$, the radio luminosity $L_R$ and the [OIII] emission line luminosity $L_{[OIII]}$.

\begin{figure}[!b]
\begin{center}
\includegraphics[height=6.6cm,width=8.4cm,angle=0]{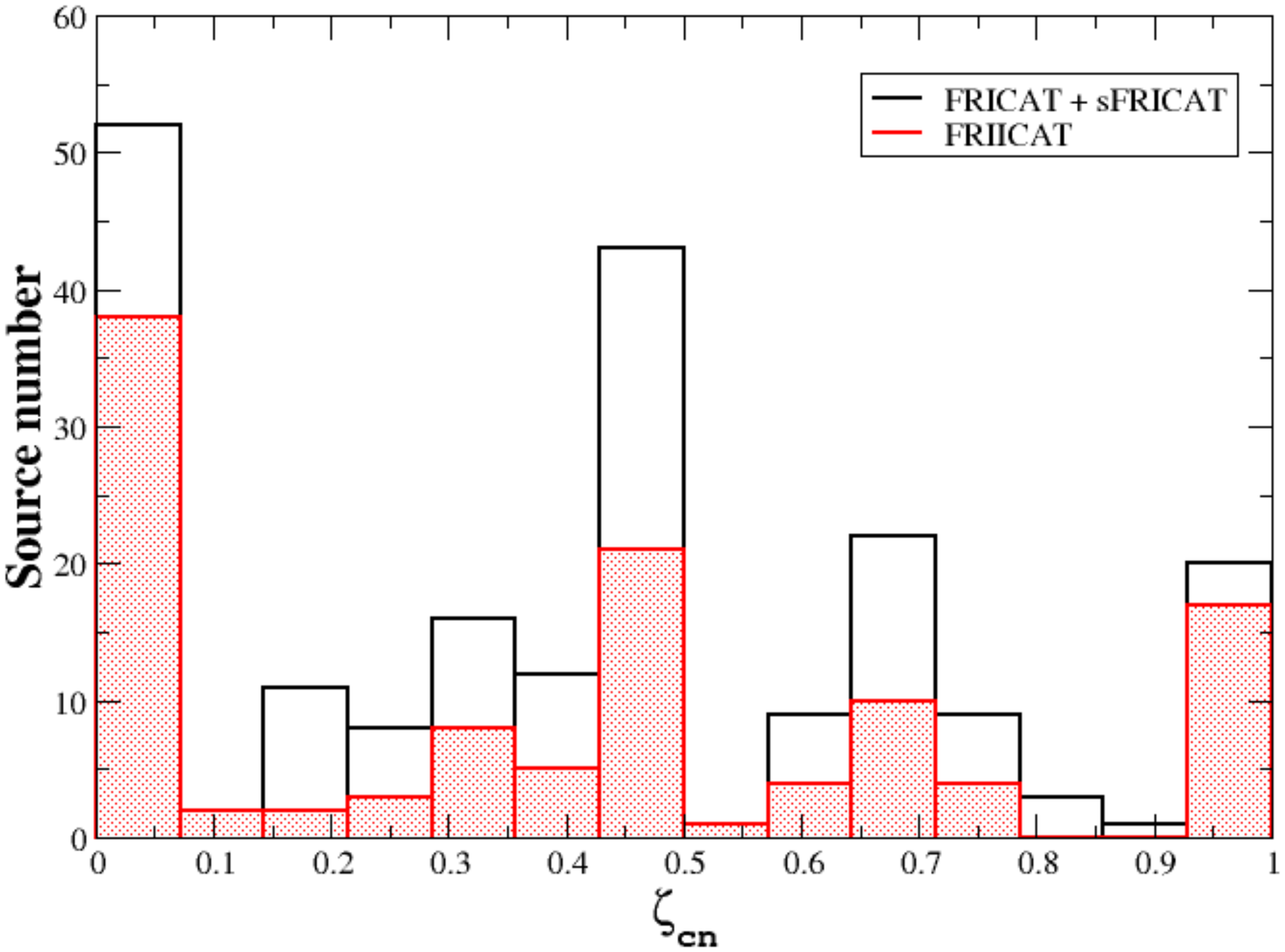}
\end{center}
\caption{The distribution of the concentration parameter $\zeta_{cn}$ estimate using the distribution of the cosmological neighbors. Black histogram refers to the FR\,I population while the red one to the FR\,IIs. The two distributions for different RG classe are similar and the largest fraction of their values lie above the value of 0.25 expected assuming a uniform distribution of cosmological neighbors. Values of $\zeta_{cn}$ in the first bin (i.e., close to zero) are simply due to the lack of cosmological neighbors within 500\,kpc distance from the central RG.}
\label{fig:ccn}
\end{figure}

The distribution of the concentration parameter $\zeta_{cn}$ for both FRICAT and FRIICAT is shown in Fig.~\ref{fig:ccn}. There are 52 FR\,Is out of 209 and 34 FR\,IIs out of a total 115, that have no cosmological neighbors within 500\,kpc distance from their position. However this does not imply that their large-scale environment is not rich since it could be due, for example, to the SDSS spectroscopic incompleteness and/or to the position of the RG in the outskirt of its galaxy-rich large-scale environment. Most of the remaining RGs indeed appear to inhabit an environment with high galaxy density. 

We note that the distribution of sources around a random position in the sky follows, in general, a uniform distribution, where the number of sources is expected to scale as $N \propto\ \vartheta^2$, with $\vartheta$ being the angular separation from the random position. Thus assuming that the same situation could be applicable to cosmological neighbors, the expected value of the $\zeta_{cn}$ is 0.25 while the largest fraction of the observed values is significantly higher.

\begin{figure}[!h]
\begin{center}
\includegraphics[height=6.6cm,width=8.4cm,angle=0]{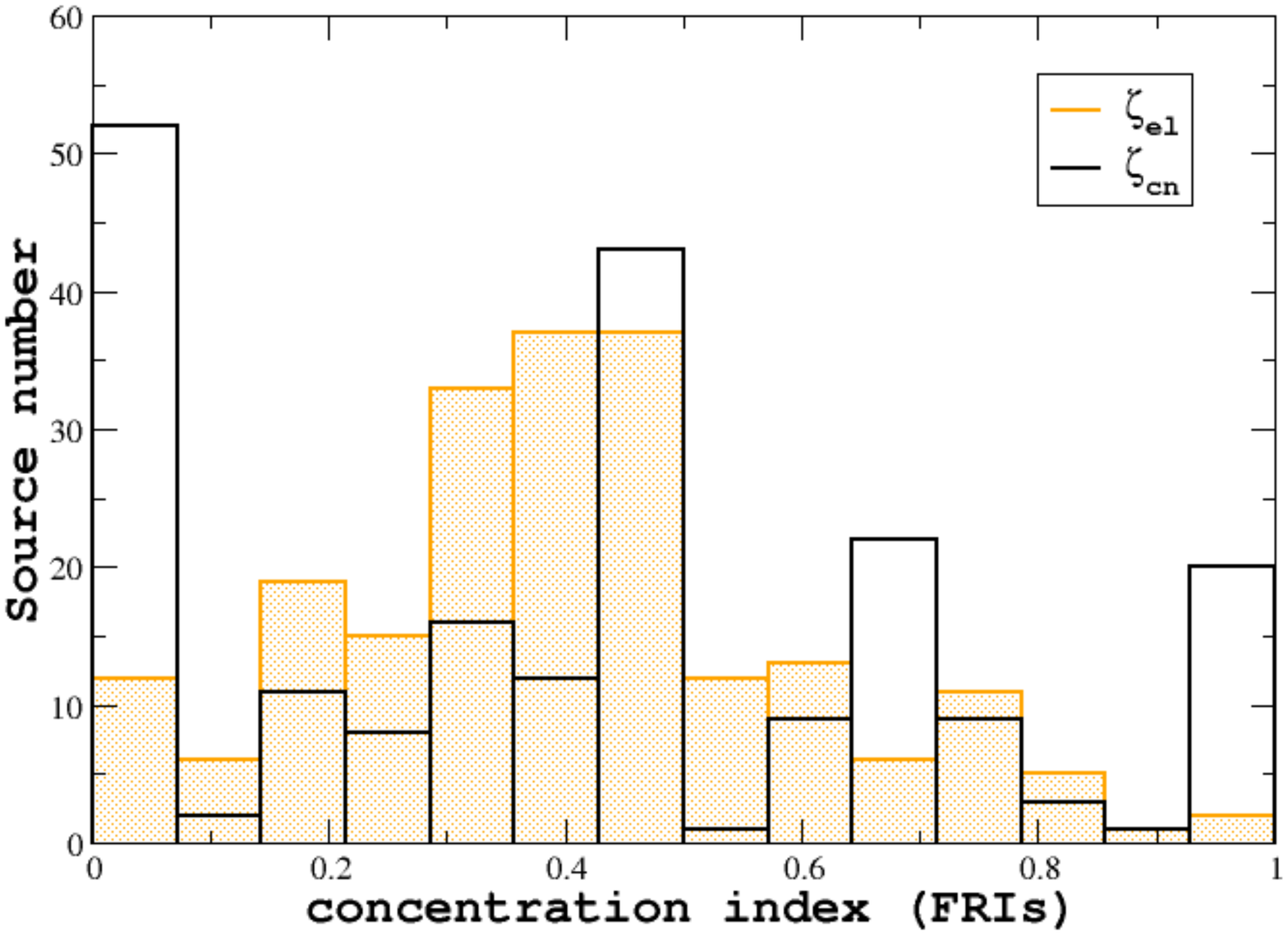}
\end{center}
\caption{The comparison between the concentration indices computed using the number of cosmological neighbors $\zeta_{cn}$ (black) and that using candidate elliptical galaxies $\zeta_{el}$ (orange) for the FR\,I population. The main difference between the two distributions is the decreasing of the low values when using $\zeta_{el}$ with respect to $\zeta_{cn}$. This is due to the fact that it is easier to find candidate elliptical galaxies with respect to cosmological neighbors. However, both parameters provide similar information.}
\label{fig:cc}
\end{figure}

In Fig.~\ref{fig:cc} we show the comparison between the concentration indices of FR\,Is computed using the number of cosmological neighbors $\zeta_{cn}$ and that of candidate elliptical galaxies $\zeta_{el}$. The distribution of $\zeta_{el}$ is peaking around 0.4 and has the largest fraction of values in the range 0.2-0.6. There are less values around zero since it is ``easier'' to find candidate elliptical galaxies than cosmological neighbors due to the SDSS fiber cladding, which prevents fibers from being placed closer than 55\arcsec. A similar situation occurs for FR\,IIs. 

We conclude that, independently of the concentration parameter adopted, RGs appear to inhabit galaxy-rich large-scale environments lying closer to the centroid of both cosmological neighbors and candidate elliptical galaxies.

\begin{figure}[!b]
\begin{center}
\includegraphics[height=6.6cm,width=8.4cm,angle=0]{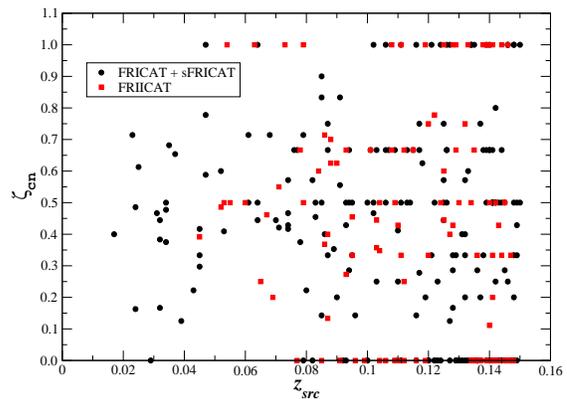}
\end{center}
\caption{The concentration parameter $\zeta_{cn}$ as function of redshift $z_{src}$ for both FR\,Is in black circles and FR\,II in red squares. No trend between these two parameters is evident thus we can claim that there is no cosmological evolution of the concentration parameter. Values of $\zeta_{cn}$ in the first bin (i.e., close to zero) are simply due to the lack of cosmological neighbors within 500\,kpc distance from the central RG.}
\label{fig:zccn}
\end{figure}

Moreover, in Fig.~\ref{fig:zccn} we also report the value of $\zeta_{cn}$ as function of the source redshift $z_{src}$. We do not see any trend/correlation between these two parameters and not even differences between the two RG classes.

\begin{figure}[!h]
\begin{center}
\includegraphics[height=6.6cm,width=8.4cm,angle=0]{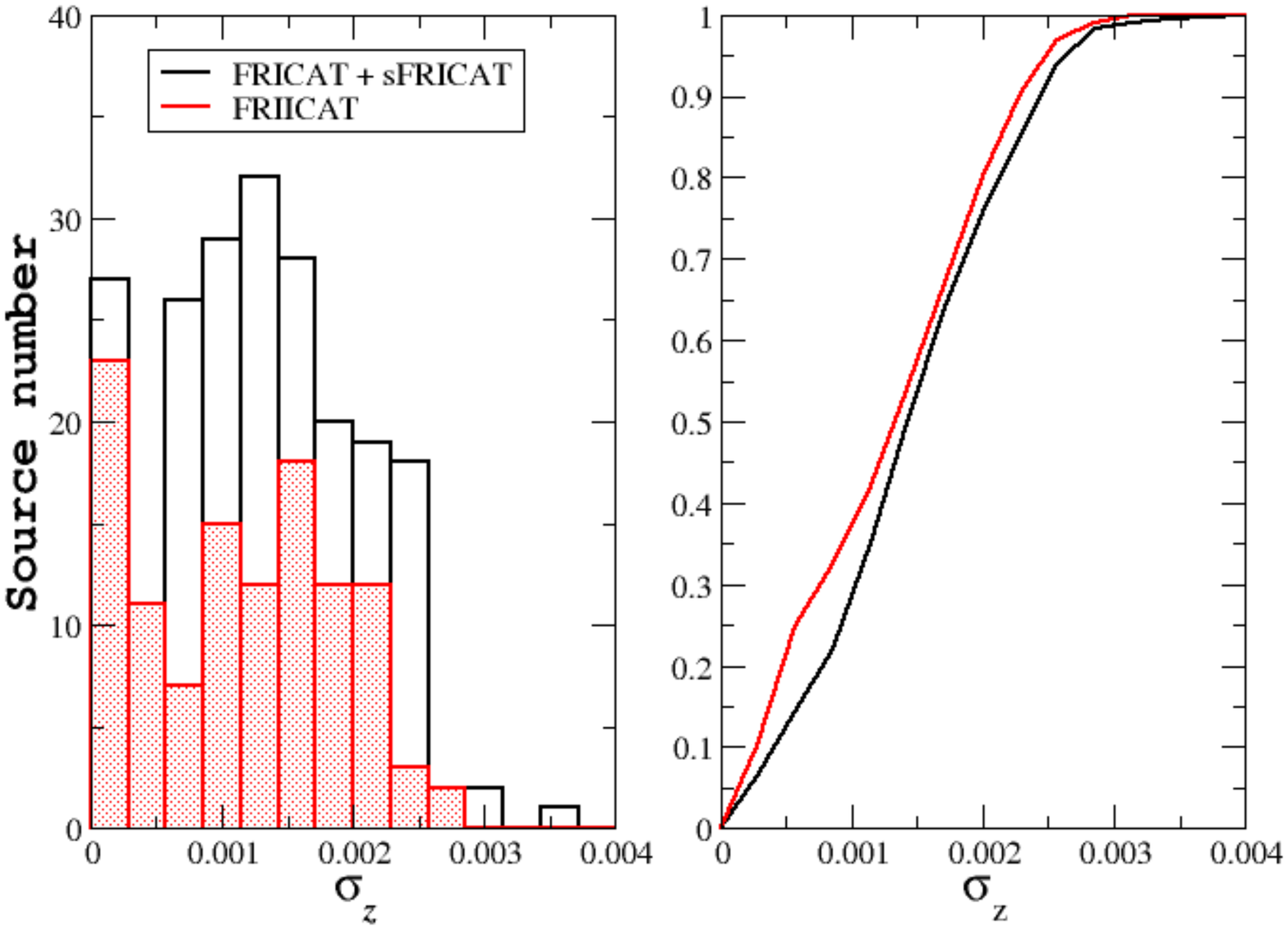}
\end{center}
\caption{In the left panel we show the histogram of the standard deviation $\sigma_{z}$ for redshift distribution of cosmological neighbors lying within 2\,Mpc. Black histogram shows the behavior of FR\,Is while the red one that of FR\,IIs. Cumulative distributions, drawn from the binned histograms, are also shown in the right panel.}
\label{fig:distsigz}
\end{figure}

\begin{figure}[!h]
\begin{center}
\includegraphics[height=6.6cm,width=8.4cm,angle=0]{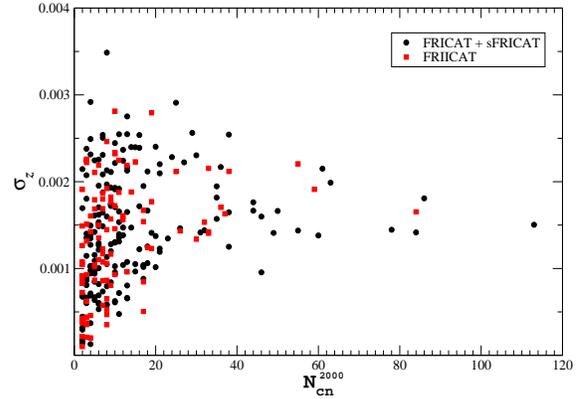}
\end{center}
\caption{The comparison between the standard deviation $\sigma_{z}$ of the redshift distribution of the cosmological neighbors and the number of cosmological neighbors within 2\,Mpc: $N_{cn}^{2000}$, a parameter used to estimate the environmental richness. There is a trend between these two parameters, with environments with larger $N_{cn}^{2000}$ appear to have larger values of $\sigma_{z}$. Black circles are referred to FR\,Is belonging to both the FRICAT and the sFRICAT while red squares mark FR\,IIs.}
\label{fig:rich}
\end{figure}

Then we explore the distribution of the average projected distance $d_m^{cn}$ of the distribution of cosmological neighbors and the standard deviation $\sigma_{z}$ of their redshift distribution. These parameters provide an estimate of the group/cluster physical size.
\begin{figure}[!h]
\begin{center}
\includegraphics[height=6.6cm,width=8.4cm,angle=0]{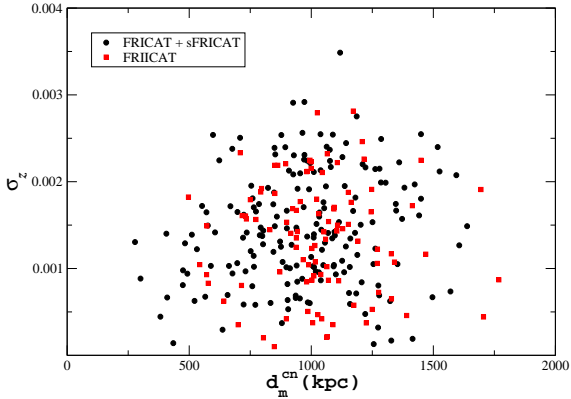}
\end{center}
\caption{The comparison between the average projected distance $d_m^{cn}$ of the distribution of cosmological neighbors and the standard deviation $\sigma_{z}$ of their redshift distribution for both RG classes, namely: FR\,Is shown as black circles while FR\,IIs as red squares. No trend is evident between these two parameters.}
\label{fig:aver}
\end{figure}
Their comparison is shown in Fig.~\ref{fig:aver}. We did not find any trend between these two parameters for both classes of RGs and as shown by their distributions in Fig.~\ref{fig:distsigz}. In the latter case, we carried out a two-sided Kolmogorov-Smirnov test and we found the $D$ parameter equal to 0.113 and corresponding to a $p$-value of 0.395 for a total number of 186 values of $\sigma_{z}$ for the FR\,Is and 96 for the FR\,IIs, { thus indicating that the  null hypothesis that the two distributions are the same cannot be rejected.}
However, as shown in Fig.~\ref{fig:rich}, we found that for high values of $N_{cn}^{2000}$ the redshift dispersion $\sigma_{z}$ appear clustered around the 0.002.

\begin{figure*}[!t]
\begin{center}
\includegraphics[height=6.6cm,width=8.4cm,angle=0]{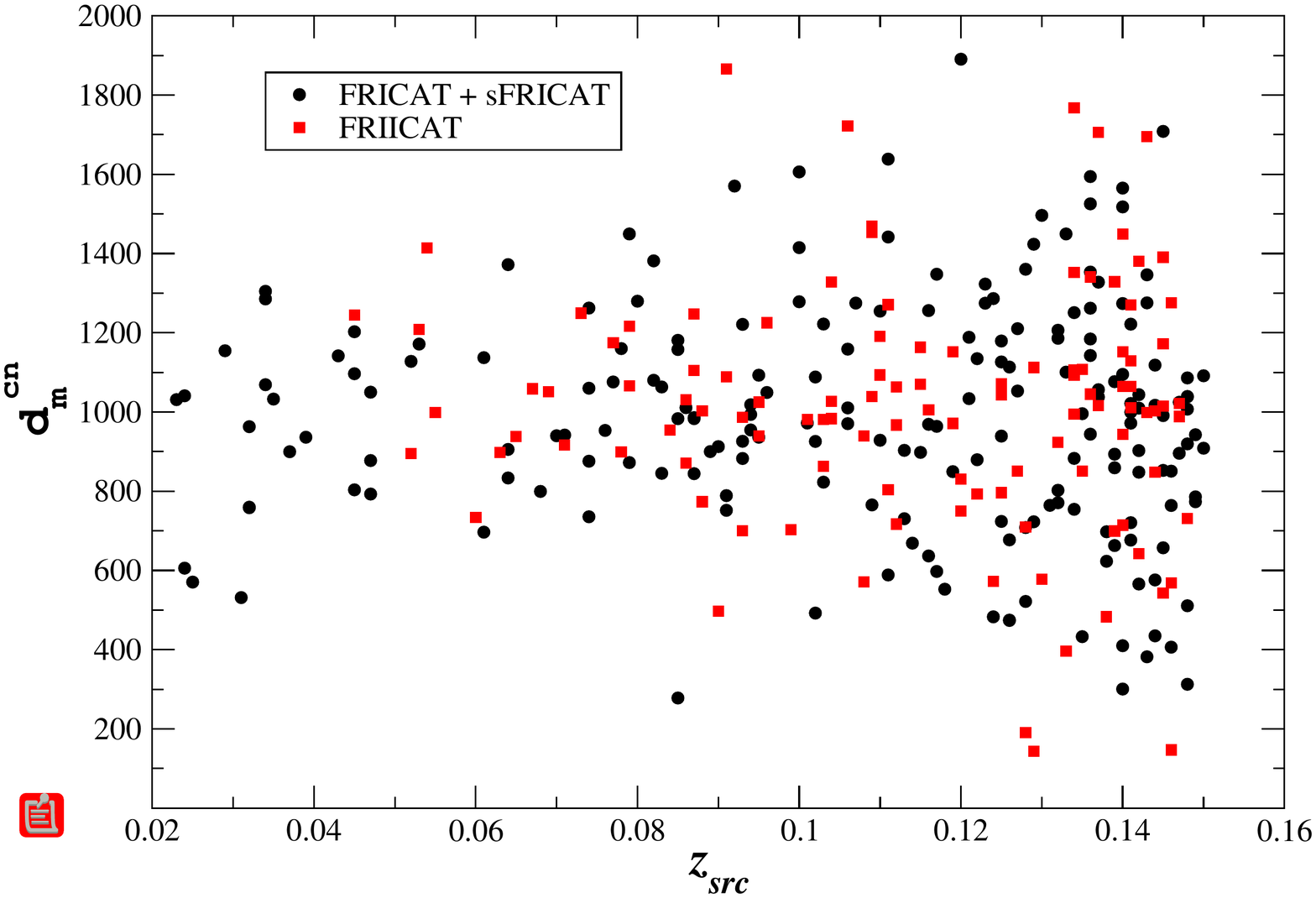}
\includegraphics[height=6.6cm,width=8.4cm,angle=0]{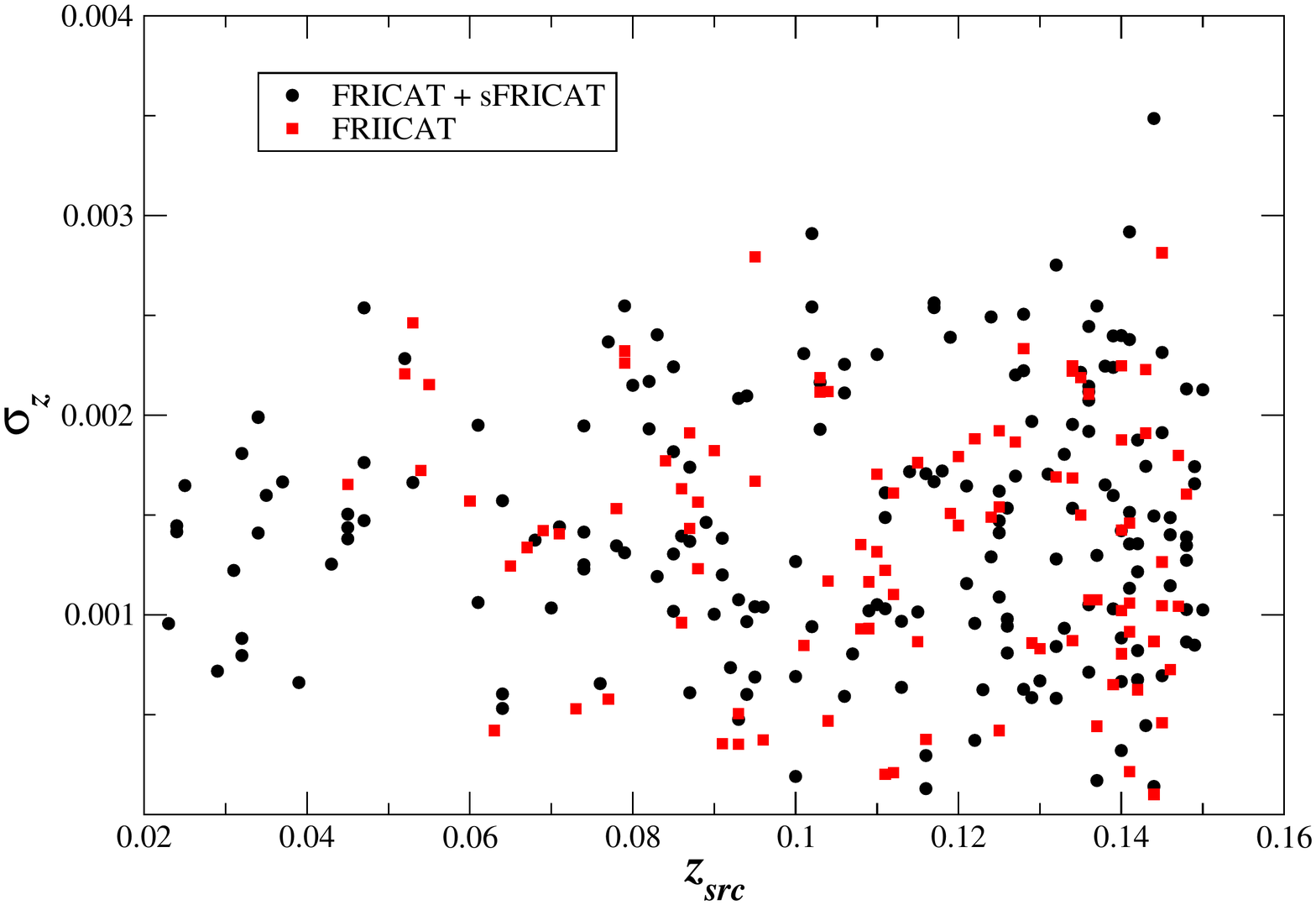}
\end{center}
\caption{Left panel) the average projected distance $d_m^{cn}$ of the distribution of cosmological neighbors vs. the redshift $z_{src}$ of the central RG. Right panel) Same of the left panel for the $\sigma_{z}$ parameter. FR\,Is are represented as black circles while FR\,IIs as red squares in both panels. No clear trend was found between these parameters and the RG redshift. }
\label{fig:zdep}
\end{figure*}

In Fig.~\ref{fig:zdep} we also show the values of both parameters, $d_m^{cn}$ and $\sigma_{z}$, as function of $z_{src}$ but no evidence of any trend is found.

Finally, we explored other links between parameters of the central RG and those of the surrounding environments but no neat trends are evident. Figures are shown in Appendix.

We show $d_m^{cn}$ and $\sigma_{z}$ in comparison with the absolute magnitude in the $R$ band of the RG where no trend/link is identified. A similar situation occurs also when comparing both these environmental parameters with radio power $L_R$ and emission line luminosity of the [OIII], i.e., $L_{[OIII]}$. The plot with $L_{[OIII]}$ allows us also to test if there are differences between LERGs and HERGs, even if given the low number of HERGs present in our sample, these results must be treated with caution making this comparison less statistically significant. We conclude that independently by the power of the central RG in different energy range or that of its host galaxy the properties of its large-scale environment is the same, being independent by the radio morphology.

\section{X-ray perspectives}
\label{sec:xray} 
\begin{figure}[]
\begin{center}
\includegraphics[height=6.6cm,width=8.4cm,angle=0]{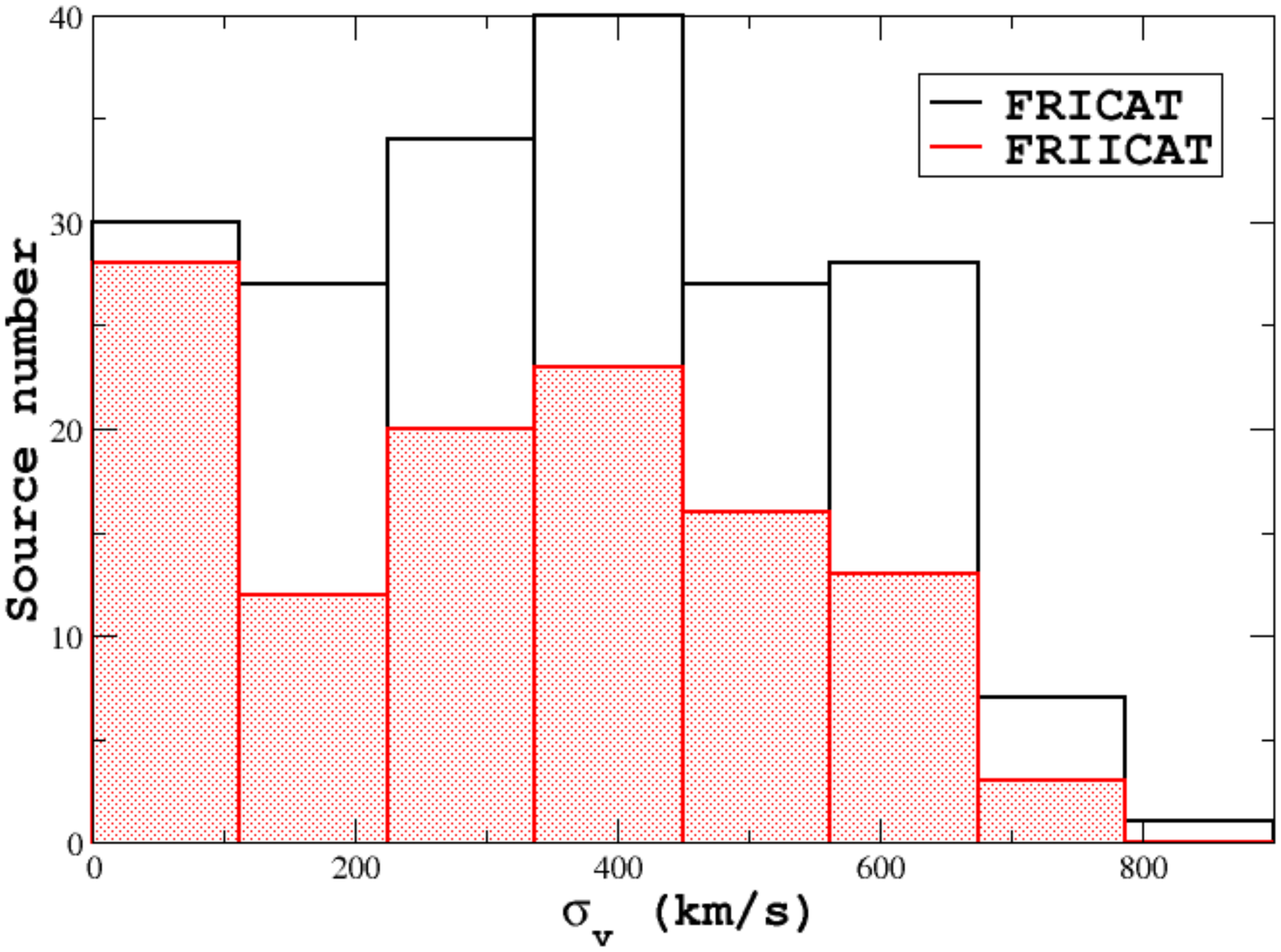}
\end{center}
\caption{The distribution of the velocity dispersion for both FRICAT and FRIICAT sources derived according to Eq. (1). As expected from their their $sigma_{z}$ distribution, they are quite similar.}
\label{fig:sigmav}
\end{figure}

We converted the standard deviation $\sigma_{z}$ of the redshift distribution computed using all cosmological neighbors within 2\,Mpc: $N_{cn}^2000$ into an estimate of the velocity dispersion $\sigma_{v}$ of galaxies in the large-scale environments of radio galaxies. We computed the standard deviation of the line-of-sight component of the velocity (i.e. radial velocity) according to Danese et al. (1980), thus using the following relation:
\begin{equation}
v_\parallel = \frac{c(z_{src}-<z>_{cn})}{1+<z>_{cn}}
\end{equation}
where $z_{src}$ is the redshift of the central radio galaxy and $\bar{z}_{cn}$ is the average redshift of all cosmological neighbors within 2\,Mpc. In Fig.~\ref{fig:sigmav} we show the distribution of the velocity dispersion computed on the basis of Eq. (1).

Then, considering $\sigma_{v}$, we also assumed that groups and clusters surrounding radio galaxies, being gravitationally bound, are virialized and thus we estimated the total mass $M_{env}$, i.e. galaxies plus IGM plus dark matter content, for their large-scale environments, according to
\begin{equation}
M_{env} = \frac{\sigma^2_{v}R_{vir}}{3G}
\end{equation}
where we assumed $R_{vir}=$2\,Mpc and we neglect the form factor 3/5.

We finally adopted two different correlations to estimate the X-ray luminosity $L_X$ and the X-ray integrated flux $F_X$. The first correlation is
\begin{equation}
L_X = 3.3\cdot10^{26} \cdot M_{env}^{1.23} erg/s
\end{equation} 
as reported by Reiprich \& B\"{o}hringer (1999) while the second directly links $\sigma_{v}$ with the X-ray luminosity: 
\begin{equation}
log \left(\frac{\sigma_{v}}{700\,km/s}\right) = -0.003 + 0.346 \cdot log \left(\frac{L_x\,E(z)^{-1}}{10^{44} erg/s}\right)
\end{equation} 
found by Clerc et al. (2016), where $E(z) = [(1+z_{src})^3*\Omega_M+\Omega_\lambda]^\frac{1}{2}$. 

\begin{figure*}[]
\begin{center}
\includegraphics[height=6.6cm,width=8.4cm,angle=0]{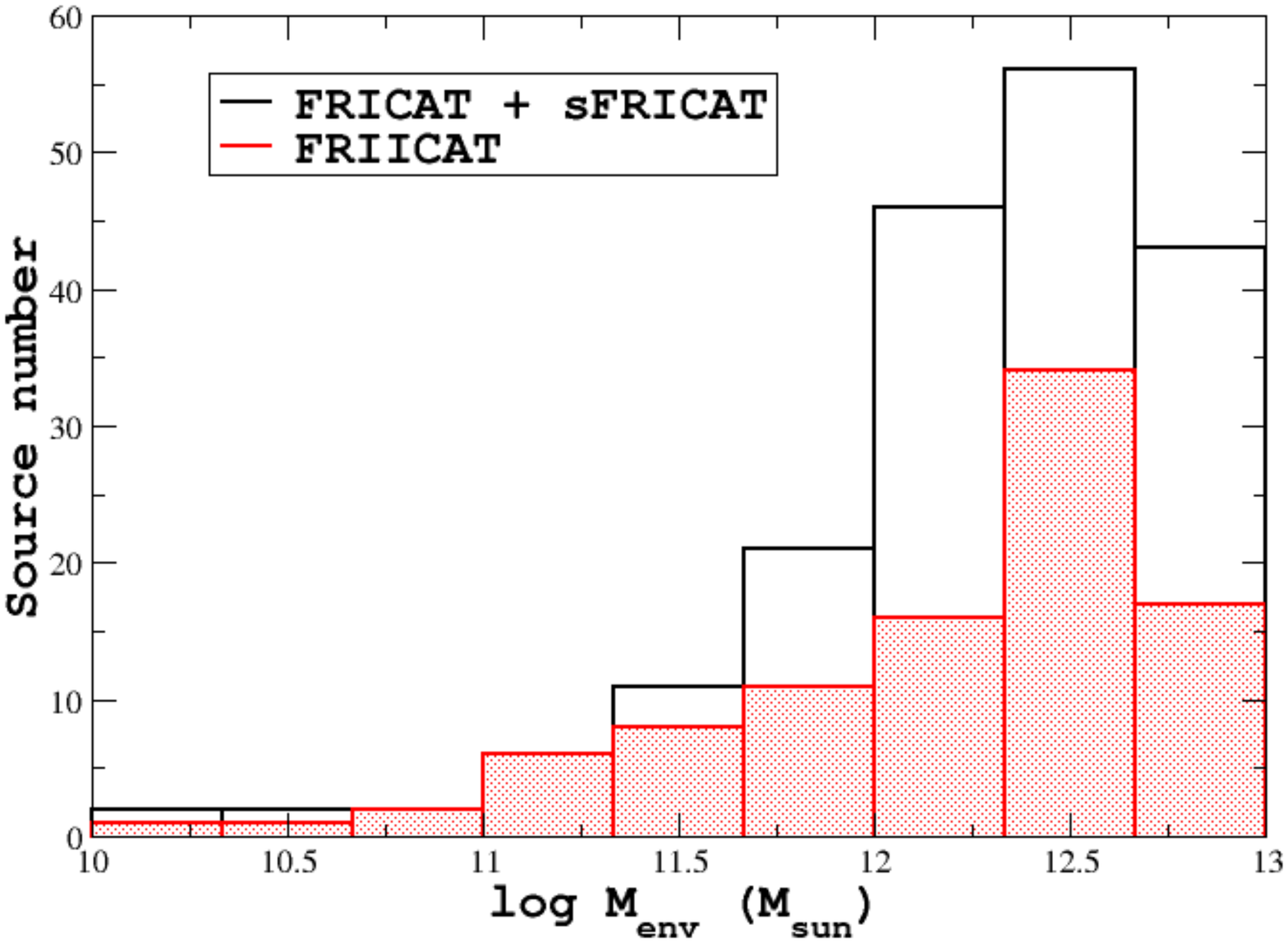}
\includegraphics[height=6.6cm,width=8.4cm,angle=0]{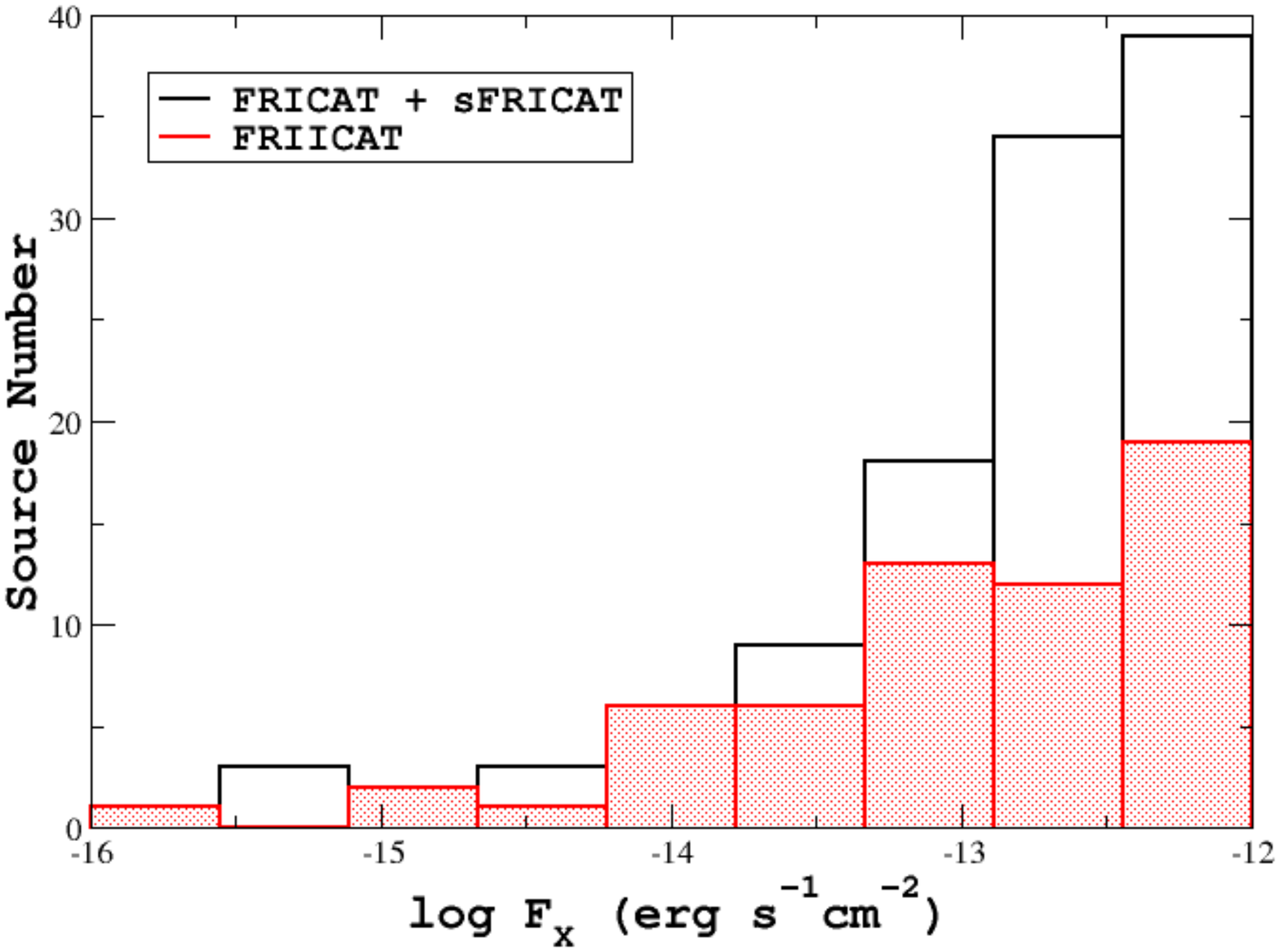}
\end{center}
\caption{Left panel) The distribution of the environmental mass derived from the virial theorem. Right panel) The distribution of the integrated X-ray fluxes estimated from the velocity dispersion and the correlation of Eq. (3).}
\label{fig:xrays}
\end{figure*}

Hence in Fig.~\ref{fig:xrays} we show the distribution of the environmental mass $M_{env}$ as well as that X-ray fluxes $F_X$ estimated according to the previous relations and, using Eq. (3), as expected, they are all below the threshold of 4.4$\times$10$^{-12}$ erg/cm/s of the ROSAT Brightest Cluster Sample \citep{ebeling98}, while the same situation occurs for more than 95\% of both radio galaxy catalogs when using Eq. (4). We remark that both $L_X$ estimates are also consistent with those derived using the $L_X$ vs $\sigma_v$ correlation found by Zhang et al. (2011).

X-ray detection of the IGM emission in the large-scale environments of these radio galaxies depends on several issues (e.g., distribution of the surface brightness, X-ray background, size of the galaxy cluster etc.). However assuming that the X-ray integrated flux is uniformly spread over a circular region of 500 kpc radius centered on the central radio source, the mean values of the X-ray intensity we estimated ranges between 3-4e-16 and 1e-14 for both FR\,Is and FR\,IIs, respectively; typical values of galaxy clusters detected by satellites as {\it XMM-Newton} and {\it Chandra} \citep[see e.g.,][]{lloyd11,mehrtens12}.

Finally, we selected 2 FR\,Is out of 6 radio galaxies for which X-ray observations are available in the \textit{XMM-Newton} archive and we carried out a simple analysis to show that low exposure times ($\sim$5 ksec) can guarantee the X-ray detection of the IGM in galaxy cluster around them as well as in the majority of the sources in explored catalogs. Selected sources are: SDSS J113359.23+490343.4 and SDSS J120401.4+201356.3 and we cut the exposure times of their archival datasets to 5.5\,ksec. 

EPIC data were retrieved from the \textit{XMM-Newton} Science Archive\footnote{http://nxsa.esac.esa.int/nxsa-web} and reduced with the \textsc{SAS}\footnote{http://www.cosmos.esa.int/web/\textit{XMM-Newton}/sas} 16.1.0 software. Following Nevalainen et al. (2005), we filtered XMM data for hard-band flares by excluding the time intervals where the high energy (\(9.5-12\) keV for MOS1 and MOS1, \(9.5-12\) keV for PN) count rate evaluated on the whole detector FOV was more than 3\(\sigma\) away from its average value. To achieve a tighter filtering of background flares, we iteratively repeated this process two more times, re-evaluating the average hard-band count-rate and excluding time intervals away more than 3\(\sigma\) from this value. The same procedure was applied to soft \(1-5\) keV band restricting the analysis to an annulus with inner and outer radii of 12' and 14', where the emission from the radio galaxy is expected to be small.

We merged data from MOS1, MOS2 and PN detectors from all observations using the \textsc{merge} task, to detect the fainter sources that would not be detected otherwise. Sources were detected on these merged images following the standard SAS sliding box task \textsc{edetect\_chain} that mainly consist of three steps: (i) source detection with local background, with a minimum detection likelihood of 8; (ii) remove sources in step 1 and create a smooth background maps by fitting a 2-D spline to the residual image; (iii) source detection with the background map produced in step 2 with a minimum detection likelihood of 10.

The task \textsc{emldetect} was then used to determine the parameters for each input source by means of a maximum likelihood fit to the input images, selecting sources with a minimum detection likelihood of 15 and a flux in the \(0.3-10\) keV band larger than \({10}^{-14}\mbox { erg}\mbox{ cm}^{-2}\mbox{ s}^{-1}\) (assuming an energy conversion factor of \(1.2\times {10}^{-11}\mbox{ cts}\mbox{ cm}^{2}\mbox{ erg}^{-1}\)). An analytical model of the point spread function (PSF) was evaluated at the source position and normalized to the source brightness. The source extent was then evaluated as the radius at which the PSF level equals half of  local background. We finally visually inspected the detected sources and removed evident spurious detections (i. e., at chip borders, in regions of diffuse emission, etc.). We then produced ``swiss-cheese" images for each detector array and observation.

\begin{figure*}[!t]
\begin{center}
\includegraphics[height=6.1cm,width=8.4cm,angle=0]{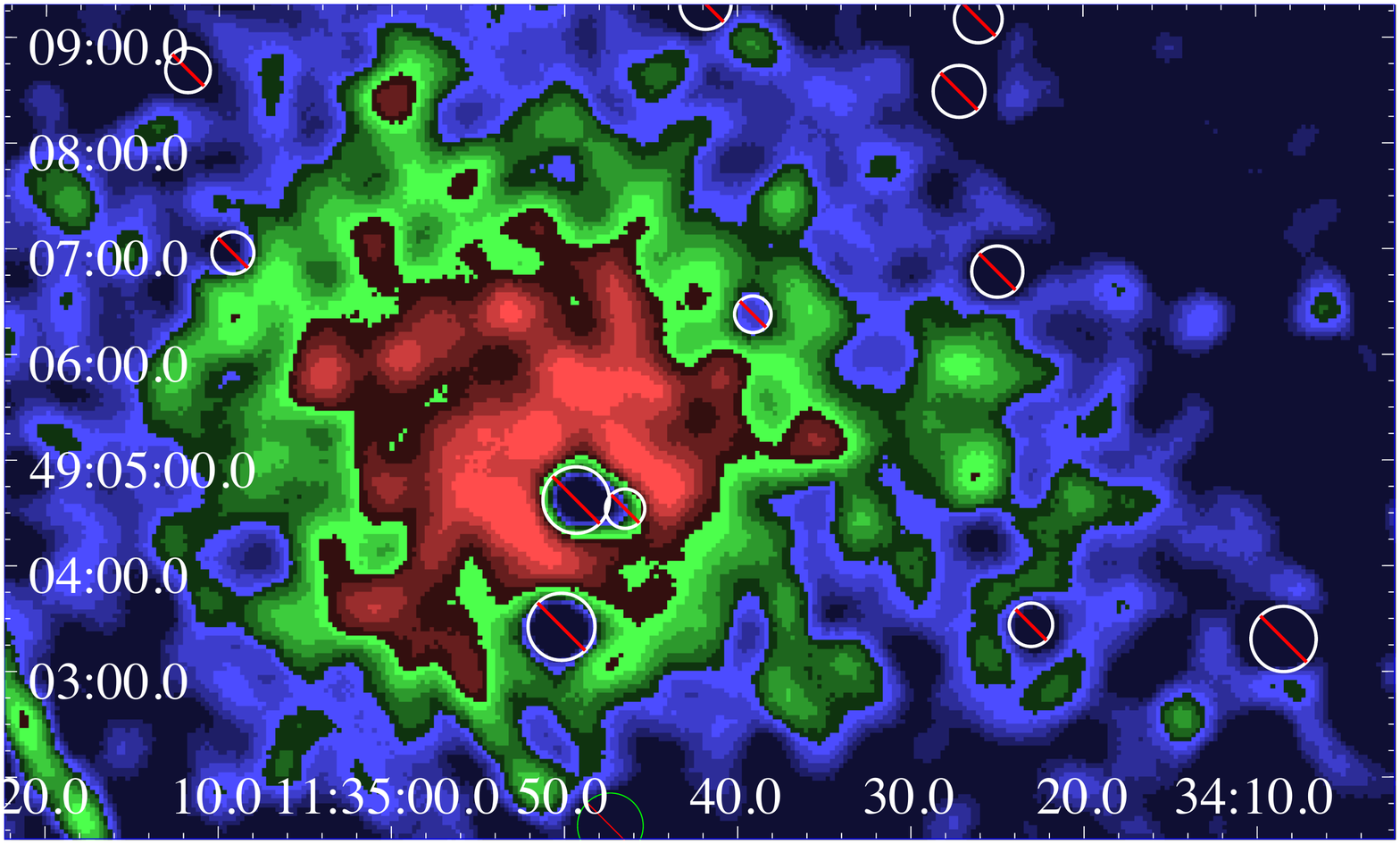}
\includegraphics[height=6.1cm,width=8.4cm,angle=0]{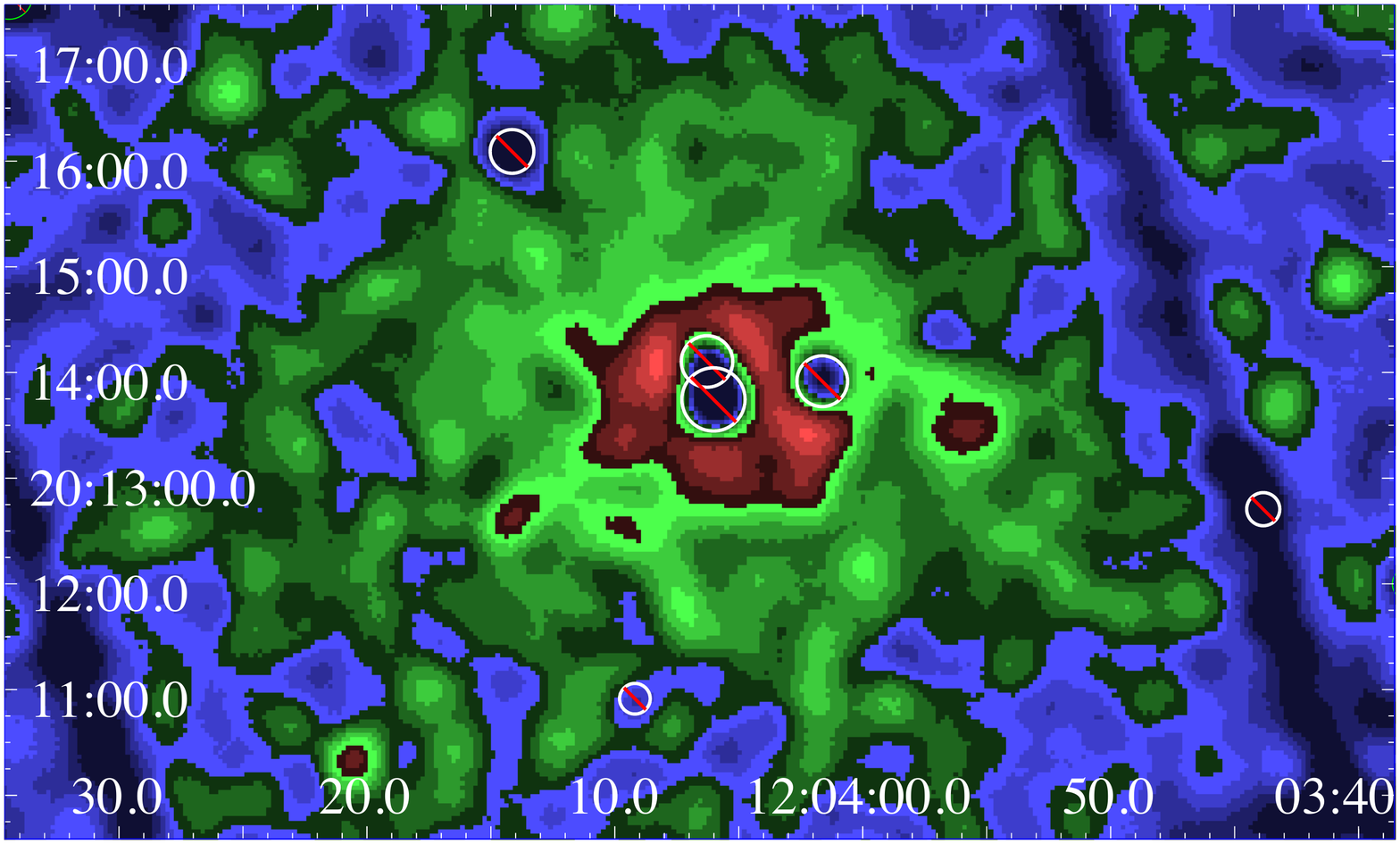}
\end{center}
\caption{The ``swiss-cheese" images for two FR\,Is having XMM-Newton archival observations cut to 5.5\,ksec exposure time, namely: SDSS J113359.23+490343.4 (left panel), SDSS J120401.4+201356.3 (right panel). These images show the clear detection of extended X-ray emission around both FR\,Is using merged XMM-Newtown observations. Images are background and point source subtracted and normalized for the exposure map. Then we also smoothed them with a Gaussian kernel of 8 pixels. White circular regions mark the location of all point sources detected during our data reduction procedure lying in the field.}
\label{fig:xrayimsb}
\end{figure*}

As shown in the upper panels of Fig.~\ref{fig:xrayimsb} we clearly detected extended X-ray emission in the merged \textit{XMM-Newton} archival images around each radio galaxy, while in the lower panels of Fig.~\ref{fig:xrayimsb} the surface brightness profiles are also reported.

It is also worth mentioning that we measured the integrated X-ray flux over a circular region of size equal to the maximum extent of the X-ray diffuse emission, evaluated by the surface brightness profiles, and it is in agreement with those values estimated using Eq. (3) and Eq.(4) for all three radio galaxies.

\section{Summary and Conclusions}
\label{sec:summary} 
We recently carried out an extensive investigation of the RG large-scale environment in the local Universe (i.e., at $z_{src} \leq$0.15). Our analysis was based on the comparison of different clustering algorithms and on the development of a new method, based on the counts of cosmological neighbors: optical sources lying within 2\,Mpc and with a redshift difference $\Delta\,z\leq$0.005 with respect to the RG lying at the center of the field examined, a method known as cosmological overdensity. Our study was also based on extremely homogeneous RG catalogs, with uniform radio, infrared and optical data available for all sources, allowing us to distinguish between FR\,I and FR\,II or LERG vs. HERG. We found that independently of their radio morphological classification they all appear to live in similar galaxy-rich large-scale environments. The same result, even if less statistically significant, was also obtained while investigating LERG vs HERG environments (M19).

Here we first performed a dedicated analysis to show how cosmological biases and artifacts can affect analyses of RG large-scale environments and how the cosmological overdensity is almost not affected. Then, even if our procedure is dependent by the SDSS selection of spectroscopic targets, it permits us to obtain better estimates of i) the ambient richness ii) the center of the galaxy distribution surrounding target RGs and iii) the redshift difference between central RG and the average values of cosmological neighbors, that seems to be a better estimate of the environmental redshift. We stress the importance of {\it comparing radio sources in the same redshift bins} to obtain a complete overview of their large-scale environments in particular for procedures not based on spectroscopic redshift estimates.

We also presented an additional statistical test based on the distribution of the median values of the RG population that showed, consistently with our previous analysis, how FR\,Is and FR\,IIs live in large-scale environments having environments that are {\it indistinguishable} on the basis of the statistical analysis we carried out, at all redshifts up to $z_{src}=$0.15. The lack of a significant number of HERGs (only 14 out of 105 total RGs) listed in the FRIICAT prevent us to draw firm conclusions on the comparison between them and LERGs and thus it was not explored.

In the current paper we also investigated several properties of the RG environments on Mpc scale using the cosmological neighbors distribution and a few parameters derived. Our main results are summarized as follows.

\begin{enumerate}
\item The concentration parameter $\zeta_{cn}$, defined as the ratio between the number of cosmological neighbors within 500\,kpc and those within 1\,Mpc, does not depend by the RG redshift $z_{src}$ and has a distribution that does not depend by the radio classification (i.e., FR\,I vs. FR\,II). Typical values of $\zeta_{cn}$ are well above 0.25 expected by a uniform distribution of surrounding cosmological neighbors. Consistent results are obtained even when using the number of candidate elliptical galaxies instead of the cosmological neighbors.
\item FR\,Is and FR\,IIs have also similar distributions of the standard deviation $\sigma_{z}$ of the redshift distribution of the cosmological neighbors as well as their average projected distance $d_m^{cn}$ computed between their position and that of the central RG. This implies that in the redshift range explored the RG environments have similar sizes.
\item When comparing the properties of the cosmological neighbors with those of the central RGs, as the absolute magnitude $M_r$, the radio power $L_R$ and $[OIII]$ emission line luminosity, we did not find any trend and any difference between FR\,I and FR\,II populations. 
\end{enumerate}

Finally, we used the $\sigma_{z}$ parameter to estimate the velocity dispersion of cosmological neighbors and assume that this is the same of all galaxies belonging to the large-scale environment of both RG populations. Then under the assumption that these RG environments are virialized we estimated the total mass $M_{env}$ present therein, i.e., mass of galaxies plus IGM plus dark matter content. 

Thanks to the correlations between $M_{env}$ and the X-ray luminosity $L_X$ of the IGM we also computed the expected values necessary to draw the feasibility of future X-ray campaigns and/or pave the path to what eROSITA will be able to detect in the near future. Given the same distribution of $\sigma_{z}$ and the lack of any trend with this parameter with those of the central radio galaxies (i.e., $M_r$, $L_R$ and $L_{[OIII]}$) the same situation occurs with the derived quantities: $M_{env}$ and $L_X$ of their large-scale environments. However, the estimates of X-ray fluxes confirm that X-ray counterpart of the groups and clusters of galaxies around FRICAT and FRIICAT sources should not to be detected in the ROSAT all sky survey demanding to {\it XMM-Newton} and {\it Chandra} facilities to get a complete view of their large-scale environments. Finally, we reduced and analyzed {\it XMM-Newton} observations of two FR\,I radio galaxies, cutting the total exposure time to 5.5 ksec, to show that a clear detection of galaxy clusters around them can be easily achieved with snapshot observations.

\appendix

\section{Figures \& Tables}
\label{app:figtab}
\begin{figure*}[]
\begin{center}
\includegraphics[height=6.6cm,width=8.4cm,angle=0]{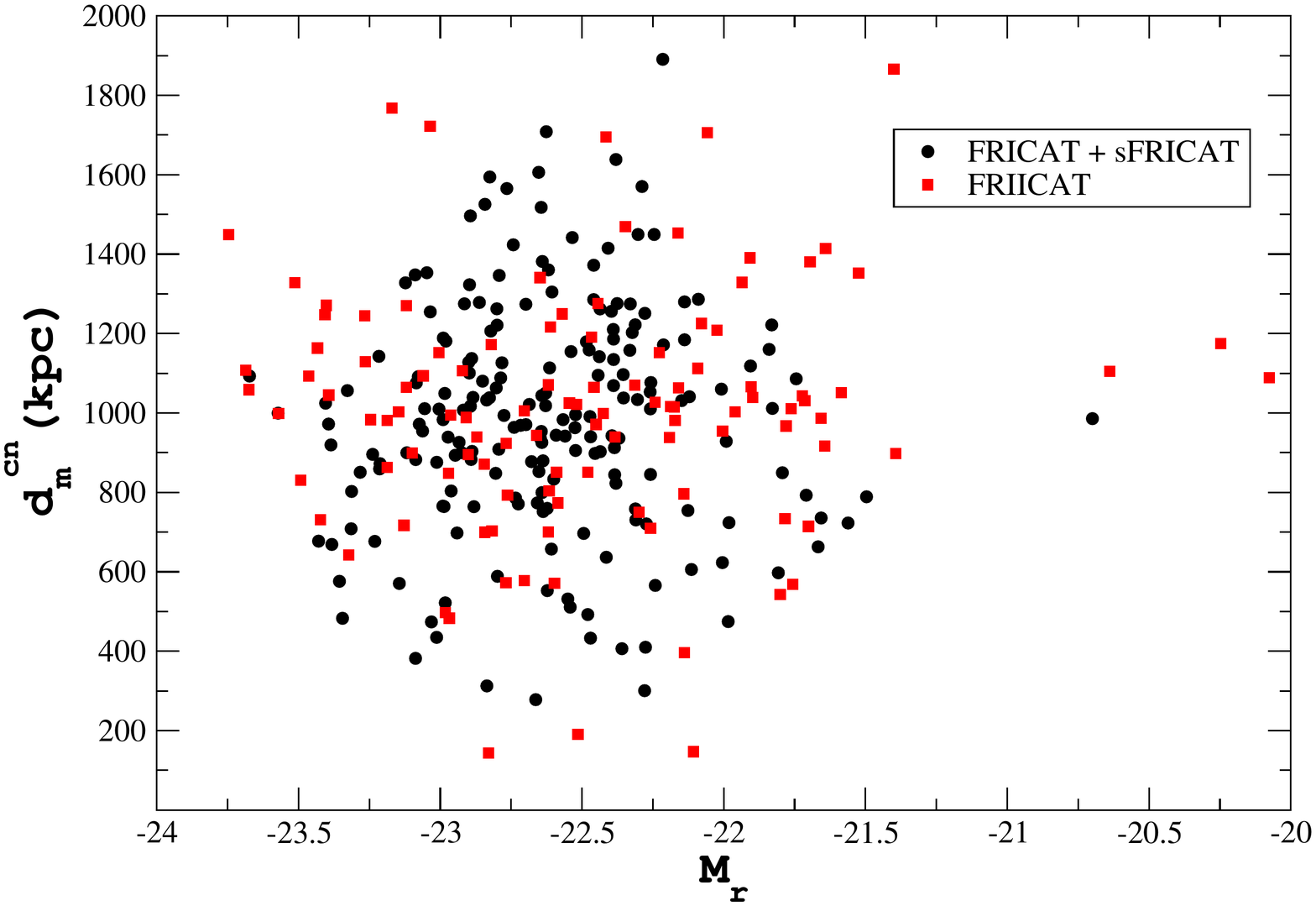}
\includegraphics[height=6.6cm,width=8.4cm,angle=0]{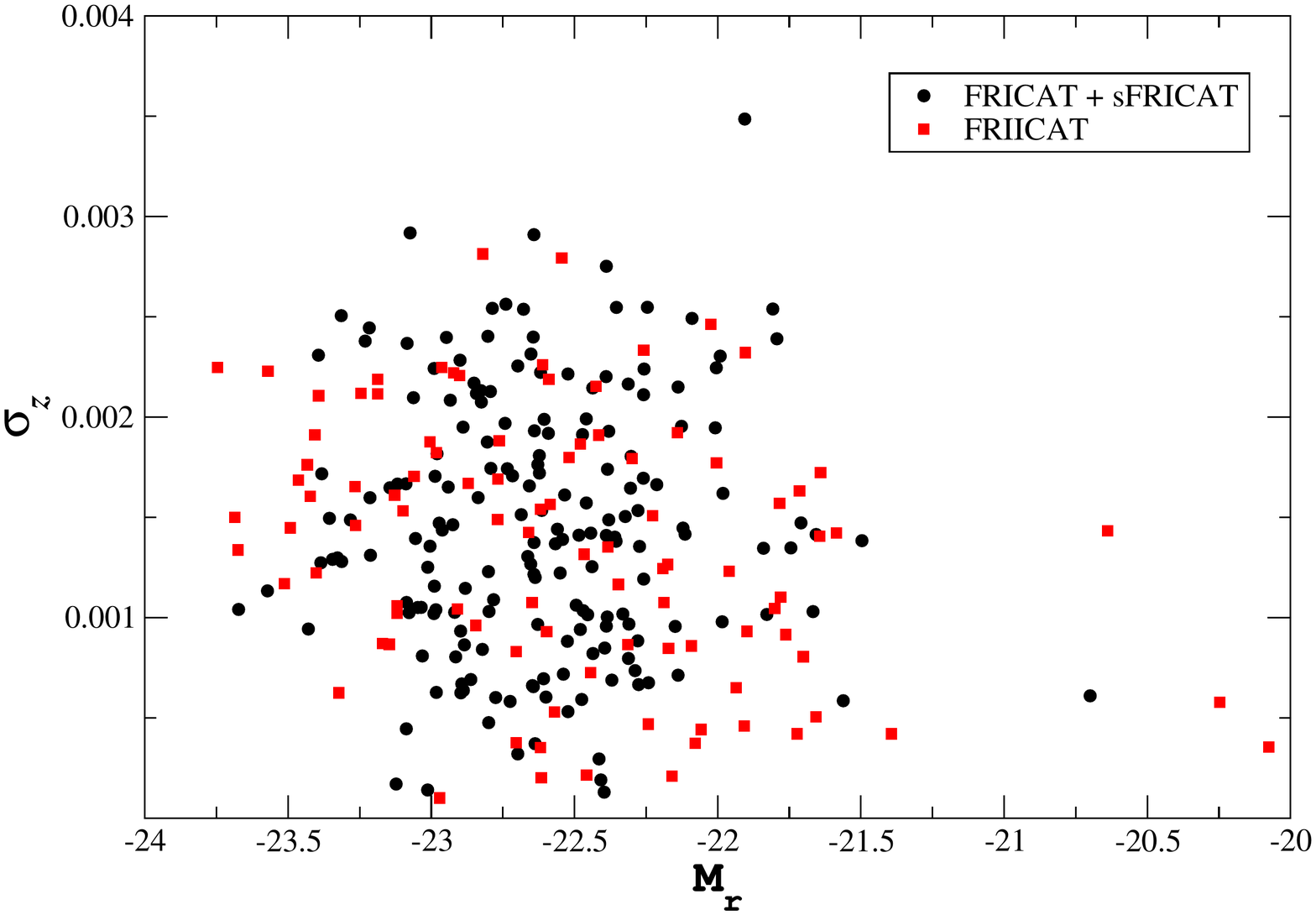}
\end{center}
\caption{Left panel) the scatter plot of the average projected distance $d_m^{cn}$ of the distribution of cosmological neighbors as function of the absolute magnitude $M_r$ of the central radio galaxy for both FR\,Is (black circles) and FR\,IIs (red squares). No neat trend or correlation is found between these two parameters. Right panel) the standard deviation $\sigma_{z}$ of redshift distribution estimated using all cosmological neighbors lying within 2\,Mpc as function of $M_r$ of the central RG. Again FR\,Is are marked as black circles while FR\,IIs as red squares.}
\label{fig:MR}
\end{figure*}

\begin{figure*}[]
\begin{center}
\includegraphics[height=6.6cm,width=8.4cm,angle=0]{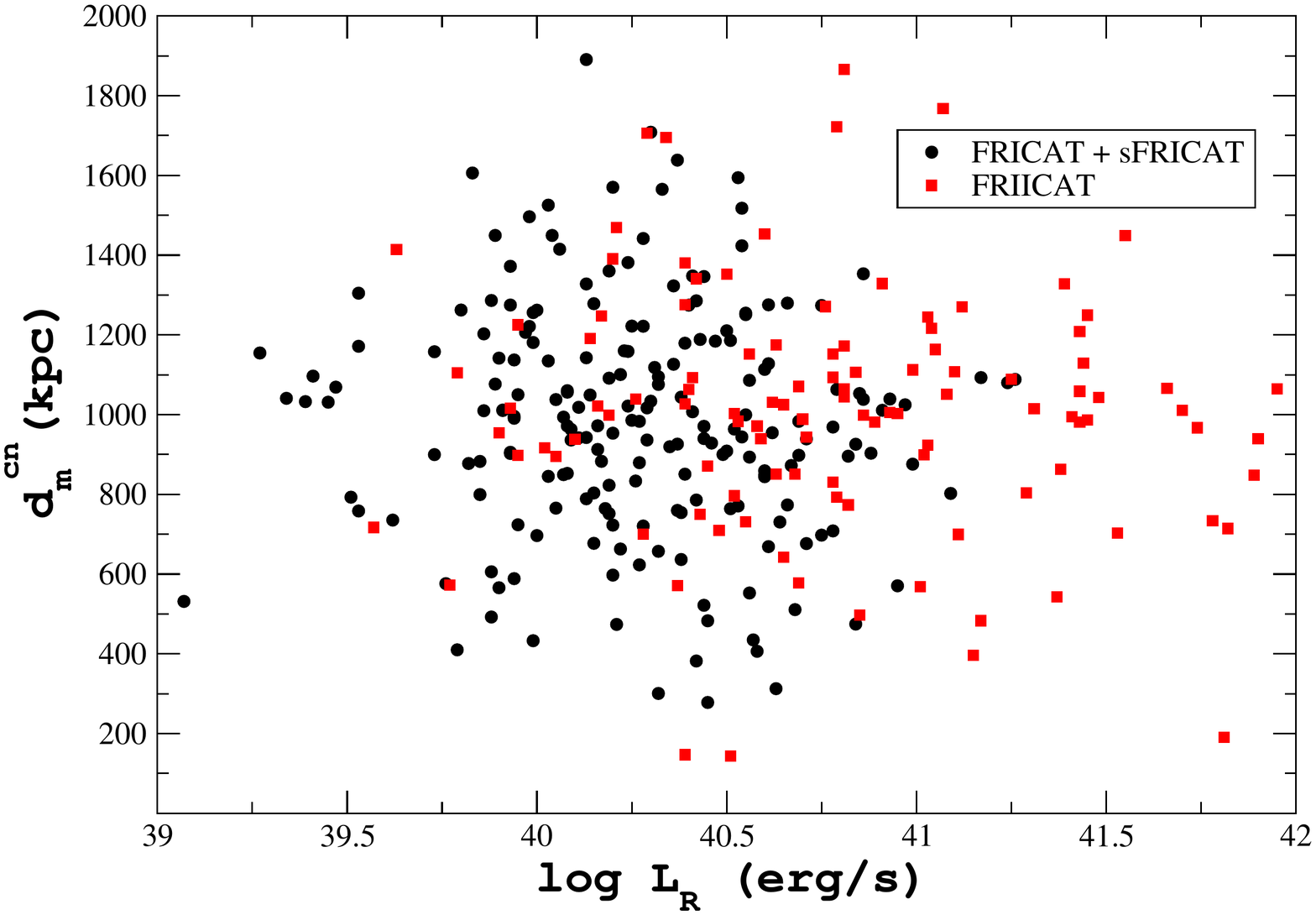}
\includegraphics[height=6.6cm,width=8.4cm,angle=0]{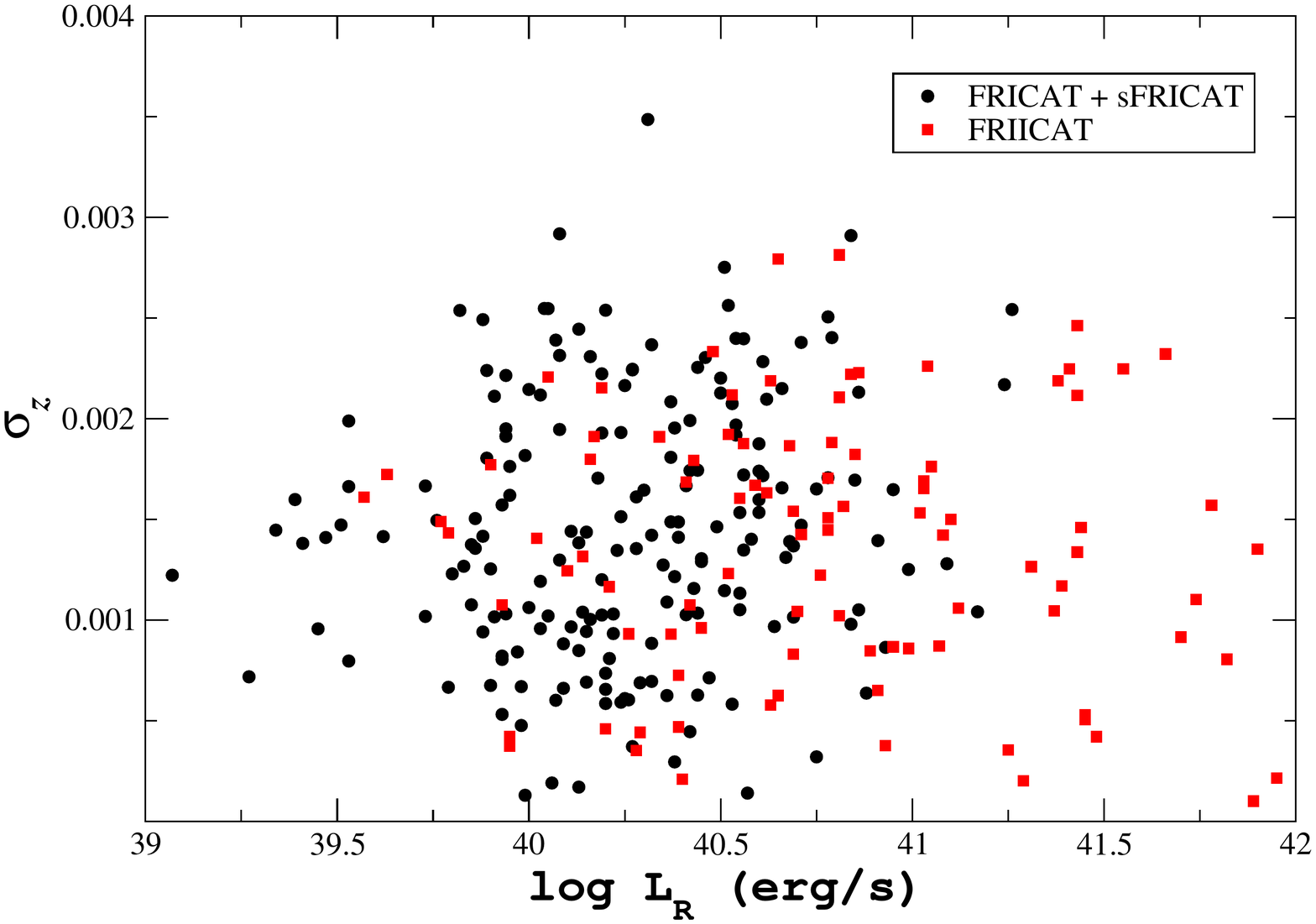}
\end{center}
\caption{Same as left panel and right panel of Fig.~\ref{fig:MR} but as function of the radio luminosity, respectively.}
\label{fig:LR}
\end{figure*}

\begin{figure*}[]
\begin{center}
\includegraphics[height=6.6cm,width=8.4cm,angle=0]{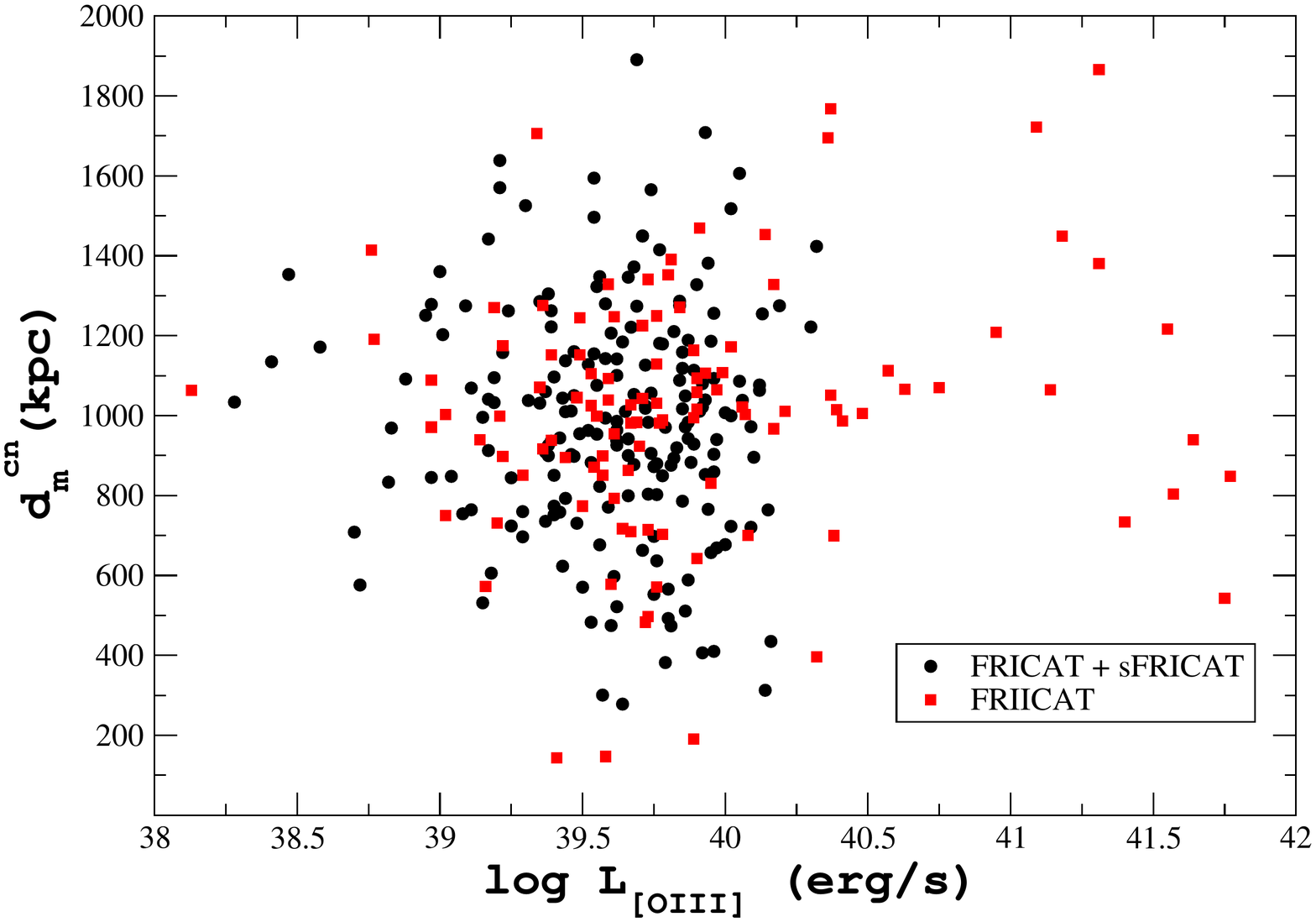}
\includegraphics[height=6.6cm,width=8.4cm,angle=0]{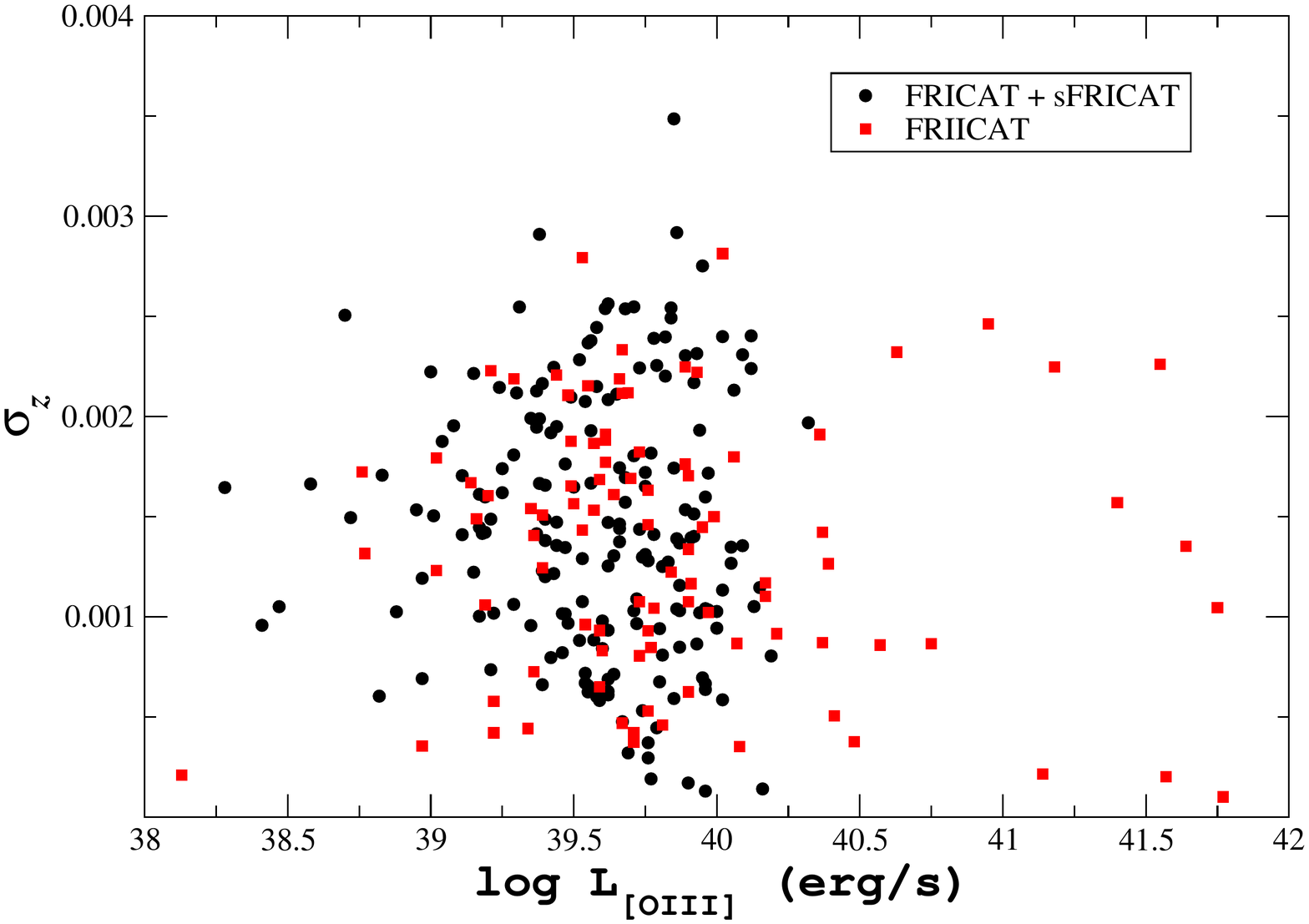}
\end{center}
\caption{Same as left panel and right panel of both Fig.~\ref{fig:MR} and Fig.~\ref{fig:LR} but as function of the [OIII] emission line luminosity, respectively.}
\label{fig:Loiii}
\end{figure*}

We did not find any trend between the environmental parameters: $d_m^{cn}$ and $\sigma_{z}$ radio galaxy parameters as $M_R$, $L_R$, $L_{[OIII]}$, but all plots are indeed shown in this Appendix A in Fig.~\ref{fig:MR}, Fig.~\ref{fig:LR} and Fig.~\ref{fig:Loiii}, respectively.
\label{app:figtab} 

We then reported here in Table~\ref{tab:main} all parameters estimated for each RG analyzed.
\begin{table*} 
\caption{Environmental parameters for the FRICAT and the FRIICAT (first 10 lines)}
\label{tab:main}
\tiny
\begin{center}
\begin{tabular}{|lllllllllllrrlll|}
\hline
SDSS  & Radio & $z_{src}$ & $\Delta\,z$ & $d_{proj}$ & $N^{500}_{cn}$ & $N^{1000}_{cn}$ & $N^{2000}_{cn}$ & $\zeta_{cn}$ & $<z_{cn}>$ & $\sigma_z$  & $d_{proj}^{cn}$ & $d_m^{cn}$ & $\sigma_v$ & $M_{env}$ & $F_X$ \\
name   & Class &                 &                   & (kpc)          &                           &                             &                             &                  &                    &                      & (kpc)         & (kpc) & (km/s) & $M_\sun$ & erg/cm$^2$s$^{-1}$ \\ 
\hline 
\noalign{\smallskip}
J073014.37+393200.4 & FRI & 0.142  & 5.3E-4 & 450.42 & 1 & 4 & 6 & 0.25 & 0.141 & 0.00082  &  207.56 &  902.48 & 215.61 & 11.86 & -13.18\\
J073505.25+415827.5 & FRI & 0.087  & 5.2E-4 & 161.92 & 3 & 4 &10 & 0.75 & 0.088 & 0.00061  &  758.00 &  985.60 & 168.04 & 11.64 & -13.05\\
J073719.18+292932.0 & FRI & 0.111  & 8.1E-4 & 159.92 & 1 & 1 & 5 & 1.00 & 0.108 & 0.00161 &  826.91 & 1441.76 & 436.15 & 12.47 & -12.07\\
J074125.85+480914.3 & FRI & 0.12   & -      & -      & 0 & 0 & 1 & -    & 0.115 & -       & 1890.54 & 1890.55 & -      & -     & -\\
J074351.25+282128.0 & FRI & 0.106  & 3.0E-4 &  63.64 & 4 & 4 & 9 & 1.00 & 0.106 & 0.00211 &  258.80 & 1010.41 & 572.42 & 12.71 & -11.69\\
J075221.83+333348.9 & FRII & 0.14  & 1.7E-4 & 111.63 & 2 & 2 & 4 & 1.00 & 0.138 & 0.00143 &  140.75 &  943.39 & 375.43 & 12.34 & -12.47\\
J075309.91+355557.1 & FRI & 0.113  & 4.5E-4 & 180.50 & 2 & 4 & 5 & 0.50 & 0.113 & 0.00097  &  229.58 &  730.53 & 260.57 & 12.02 & -12.73\\
J075506.67+262115.9 & FRI & 0.123  & -      & -      & 0 & 0 & 1 & -    & 0.123 & -       & 1274.66 & 1274.66 & -      & -     & -\\
J075529.95+520450.6 & FRII & 0.14  & -      & -      & 0 & 0 & 0 & -    & -     & -       & -       & -       & -      & -     & -\\
J075628.78+501716.3 & FRII & 0.134 & -      & -      & 0 & 0 & 2 & -    & 0.133 & 0.00087  &  322.70 & 1768.24 & 230.16 & 11.91 & -13.04\\
\noalign{\smallskip}
\hline
\end{tabular}\\
\end{center}
Col. (1): SDSS name. \\
Col. (2): Radio class to distinguish between source belonging to the FRICAT and the FRIICAT.\\
Col. (3): source redshift. \\
Col. (4): Absolute value of the redshift difference between the RG and the closest galaxy cluster in the T12 catalog.\\
Col. (5): Physical distance between the RG and the closest galaxy cluster in the T12 catalog. This is computed at the $z_{src}$ of the central RG.\\
Col. (6,7,8): Number of cosmological neighbors within 500, 1000 ad 2000 kpc, respectively, estimated at the $z_{src}$ of the central radio galaxy.\\
Col. (9): The concentration parameter $\zeta_{cn}$.\\
Col. (10): The average redshift of the cosmological neighbors in 2\,Mpc.\\
Col. (11): The standard deviation of the redshift distribution for the cosmological neighbors within 2\,Mpc.\\
Col. (12): Physical distance between the central RG and the average position of the cosmological neighbors within 2\,Mpc. This is computed at the $z_{src}$ of the central RG.\\
Col. (13): Average distance of the cosmological neighbors within 2\,Mpc.\\
Col. (14): Velocity dispersion of the cosmological neighbors located in 2\,Mpc.\\
Col. (15): Environmental mass estimated from the velocity dispersion.\\
Col. (16): X-ray flux estimated from the velocity dispersion.\\
\end{table*}

\acknowledgments
We thank the anonymous referee for useful and valuable comments that led to improvements in the paper. In particular, all her/his statistical notes that have been implemented in the revised version of the manuscript.
F. M. wishes to thank Dr. C. C. Cheung for their valuable discussions on this project initially planned during the IAU 313 on the Galapagos islands. 
This work is supported by the ``Departments of Excellence 2018 - 2022'' Grant awarded by the Italian Ministry of Education, University and Research (MIUR) (L. 232/2016). This research has made use of resources provided by the Compagnia di San Paolo for the grant awarded on the BLENV project (S1618\_L1\_MASF\_01) and by the Ministry of Education, Universities and Research for the grant MASF\_FFABR\_17\_01. This investigation is supported by the National Aeronautics and Space Administration (NASA) grants GO6-17081X and GO9-20083X. F.M. acknowledges financial contribution from the agreement ASI-INAF n.2017-14-H.0.
Funding for SDSS and SDSS-II has been provided by the Alfred P. Sloan Foundation, the Participating Institutions, the National Science Foundation, the U.S. Department of Energy, the National Aeronautics and Space Administration, the Japanese Monbukagakusho, the Max Planck Society, and the Higher Education Funding Council for England. The SDSS Web Site is http://www.sdss.org/. The SDSS is managed by the Astrophysical Research Consortium for the Participating Institutions. The Participating Institutions are the American Museum of Natural History, Astrophysical Institute Potsdam, University of Basel, University of Cam- bridge, Case Western Reserve University, University of Chicago, Drexel University, Fermilab, the Institute for Advanced Study, the Japan Participation Group, Johns Hopkins University, the Joint Institute for Nuclear Astrophysics, the Kavli Institute for Particle Astrophysics and Cosmology, the Korean Scientist Group, the Chinese Academy of Sciences (LAMOST), Los Alamos National Laboratory, the Max- Planck-Institute for Astronomy (MPIA), the Max-Planck- Institute for Astrophysics (MPA), New Mexico State University, Ohio State University, University of Pittsburgh, University of Portsmouth, Princeton University, the United States Naval Observatory, and the University of Washington. TOPCAT and STILTS astronomical software \citep{taylor05} were used for the preparation and manipulation of the tabular data and the images.

~

\end{document}